\documentclass[11pt,en,cite=authoryear]{elegantpaper}

\title{Preliminary experimental results of determining the geopotential difference between two synchronized portable hydrogen clocks at different locations}
\author{ Wen-Bin Shen$^{1,2,*}$ \and Kuangchao Wu$^1$ \and Xiao Sun$^1$ \and Chenghui Cai$^1$ \and Ziyu Shen$^3$}
\institute{$^1$ Department of Geophysics, School of Geodesy and Geomatics/Key Laboratory of Geospace Environment and Geodesy of Ministry of Education, Wuhan University, Wuhan 430079, China \\
$^2$ State Key Laboratory of Information Engineering in Surveying, Mapping and Remote Sensing, Wuhan University, Wuhan 430079, China \\
$^3$ School of Resource and Environment, Hubei University of Science and Technology, Xianning, Hubei, China \\
$^*$ Corresponding author: wbshen@sgg.whu.edu.cn}

\usepackage{array}  
\usepackage{amsmath,bm}     
\usepackage{subfigure}

\usepackage{amssymb}
\usepackage{verbatim}
\usepackage{indentfirst}
\usepackage{multirow}
\usepackage{geometry}
\usepackage{hyperref}
\usepackage{times}

\begin{document}

\maketitle

\begin{abstract}
According to general relativity theory (GRT), by comparing time elapses between two precise clocks located at two different stations, the gravity potential (geopotential) difference between them can be determined,  due to the fact that precise clocks at positions with different geopotentials  run at different rates. Here, we provide preliminary experimental results of the geopotential determination based on time elapse comparisons between two remote atomic clocks located at Beijing and Wuhan, respectively. After synchronizing two hydrogen atomic clocks at Beijing 203 Institute Laboratory (BIL) for 20 days as zero-baseline calibration, namely synchronization, we transport one clock to Luojiashan Time-Frequency Station (LTS),  Wuhan,  without stopping its running.  Continuous comparisons between the two remote clocks were conducted for 65 days based on the Common View Satellite Time Transfer (CVSTT) technique. The ensemble empirical mode decomposition (EEMD) technique is applied to removing the uninteresting periodic signals contaminated in the original CVSTT observations to obtain the residual clocks-offsets series, from which the time elapse between the two remote clocks was determined. Based on the accumulated time elapse between these two clocks the geopotential difference between these two stations was determined. Given the orthometric height (OH) of BIL, the OH of the LTS was determined based on the determined geopotential difference. Comparisons show that the OH of the LTS determined by time elapse comparisons deviates from that determined by Earth gravity model EGM2008 by about 98 m. The results are consistent with the frequency stabilities of the hydrogen atomic clocks (at the level of $10^{_{-15}}$/day) applied in our experiments.  Here the EEMD technique was first introduced and applied in the subjects related to this study, especially for the purpose of removing the periodic signals from the original CVSTT observations to obtain the residual clock offsets signals, from which one can more effectively extract the geopotential-related signals. In addition, we used 85-days original observations to determine the geopotential difference between two remote stations based on the CVSTT technique.  Using more precise atomic or optical clocks, the CVSTT method for geopotential determination could be applied effectively and extensively in geodesy in the future.
\end{abstract}
	

\section{Introduction}
\label{sec:1}
Precise determination of the Earth's geopotential field and orthometric height (OH) is a main task in geodesy \citep{Mazurova2012Comparison, Mazurova2017Development}. With rapid development of time and frequency science, high-precision atomic clock manufacturing technology \citep{Hinkley2013An, Bloom2014, McGrew2018} provides an alternative way to precisely determine the geopotentials and OHs, which has been extensively discussed in recent years \citep{Bondarescu2012Geophysical, Shen2016Formulation, Lion2017Determination, Shen2018Formulation}, and opening a new era of time-frequency geoscience \citep{Kopeikin2018Normal, SHEN2019}.

General relativity theory (GRT) states that a precise clock runs at different rates at different positions with different geopotentials \citep{Einstein1915}. Based on GRT, \cite{Bjerhammar1985On} proposed an idea to determine the geopotentials via clock transportation. Later, \cite{Shen1993n} put forward an approach to determine geopotentials via gravity frequency shift (GFS) that is also based on GRT. Consequently, the geopotential difference between two arbitrary points can be determined by comparing the running rates of two precise atomic clocks \citep{Bjerhammar1985On, Mai2013} or by comparing their frequencies \citep{Shen1993n, Shen1998, Shen2009, Shen2011n}. In order to determine geopotentials with an accuracy of  0.1 m$^{2}$s$^{-2}$ (equivalent to 1 cm in OH), the atomic clocks with frequency stabilities of  $1\times 10^{-18}$ are required.

High performance clocks and time-frequency transfer techniques have been intensively developing over the past 60 years \citep{Akatsuka2008Optical, Bondarescu2015The, Mehlst2018Atomic}. Recently, optical-atomic clocks (OACs) with stabilities around $ 10^{-18}$ level have been successively generated \citep{T2015Systematic,Campbell2017A, McGrew2018}, and in the near future mobile high-precision satellite-borne optical clocks will be in practical usage \citep{Poli2014A, Singh2015Development, Riehle2017Optical}. This provides a good opportunity to precisely determine the geopotentials via precise clocks. Therefore, high-precision clocks enable 1 cm level geopotential determination and  the realization of world height system (WHS) unification \citep{Shen2016Formulation, Koller2016A,M2018High}.

As mentioned before, there are two kinds of methods for determining the geopotentials via clocks. One method is to compare the time elapses of two clocks located at two stations using various techniques \citep{Bjerhammar1985On, Shen2015Geopotential, Kopeikin2016Chronometric}. The other method is to compare frequency difference of the two clocks \citep{Shen2017Determination, Shen2018Formulation, Deschenes2015Synchronization, Weiss2017A}.  It is true that there exist some successful transportation experiments for determining  geopotential via high-precise fiber links \citep{Lisdat2016A,Takano2016, McGrew2018, Geodesy2018}, and the corresponding results are quite successful. For example, \cite{Geodesy2018} provided the transportation experiments by using optical atomic clocks for their unprecedented stability, and the discrepancy between relativistic method and conventional method variates 2 $\thicksim$ 42 $m^{2}/s^{2}$ in geopotential units. However, there are seldom transportation experiments using the satellite signals for comparing clock offsets in the relativistic geodesy. In fact, satellite time and frequency signal transfer approach is most prospective in the future due to the fact that it is not constrained by geographic conditions, for instance connecting two continents separated by oceans.  \cite{Kopeikin2016Chronometric} discussed and provided experimental results of transportation experiments for geopotential difference determination via Common-View Satellite Time Transfer (CVSTT)  technique, with the help of two hydrogen atomic clocks. The results are consistent with the  clocks' stabilities in their experiments with a discrepancy of hundreds of meters. However, in their experiments, the time comparison period is quite short. For instance, the geopotential difference measurement lasted about 63900 s (17.7 hr), and the zero-baseline calibration lasted only 23790 s (6.6 hr). 

In this study, we focus on the clocks' time elapse comparisons via the CVSTT technique, and  the determination of geopotential difference  based on clock transportation experiments. Here, the transportation experiments last for a quite long time, with the zero-baseline measurement lasting for 20 days, and the geopotential difference measurement lasting for 65 days with plenty of data sets. In addition, we first use the EEMD technique to remove the uninterested periodic signals from the original CVSTT observations to obtain the residual clock offsets signals, and then extract the geopotential-related signals from the residual clock offsets signals. In our knowledge, this is a first application in CVSTT data processing for the purpose of determining the geopotential difference in geodetic community. In this study, the stability of H-masers used in the experiments is at $10^{-15}$/day level, which means that the determination of geopotential difference is limited to tens of meters in equivalent height. We admit that the performance of the H-masers cannot be compared to the optical atomic clocks with unprecedented stability, and it cannot obtain perfect results like the optical atomic clocks via optical fiber. However, the H-masers have their special advantages, for example, they can keep operation with a relative stable stability for a long period, which could be difficult for the optical atomic clocks at the current stage. In the future, the optical atomic clock might be introduced in the similar experiments when the experimental conditions are mature.

This paper is organized as follows. In Section 2, we briefly describe how to determine geopotential as well as OH using two remote clocks. In Section 3, we focus on describing how to achieve the comparisons of time elapses between two remote clocks based on the CVSTT technique, and how to cancel or greatly reduce various error sources. In Section 4, we introduce ensemble empirical mode decomposition (EEMD) technique, and verify its effectiveness for extracting the linear signals of interest from the original CVSTT observations based on simulation experiments. In Section 5, we describe our experimental setup, records, and data procession. We provide experimental results and relevant discussions in Section 6, and draw conclusions in Section 7.


\section{Method}
\label{sec:2}
Based on GRT, considering the proper time durations $\Delta t_{_A}$ and $\Delta t_{_B}$  recorded by two precise clocks $C_{A}$ and $C_{B}$ located at two stations $A$ and $B$, respectively (see Fig. \ref{fig:1}). Accurate to $1/c^{2}$, the following equation holds \citep{Weinberg1972General, Bjerhammar1985On, Shen2009An, Mehlst2018Atomic}:

\begin{figure}[hbt]
	\centering
	\includegraphics[width=1\textwidth]{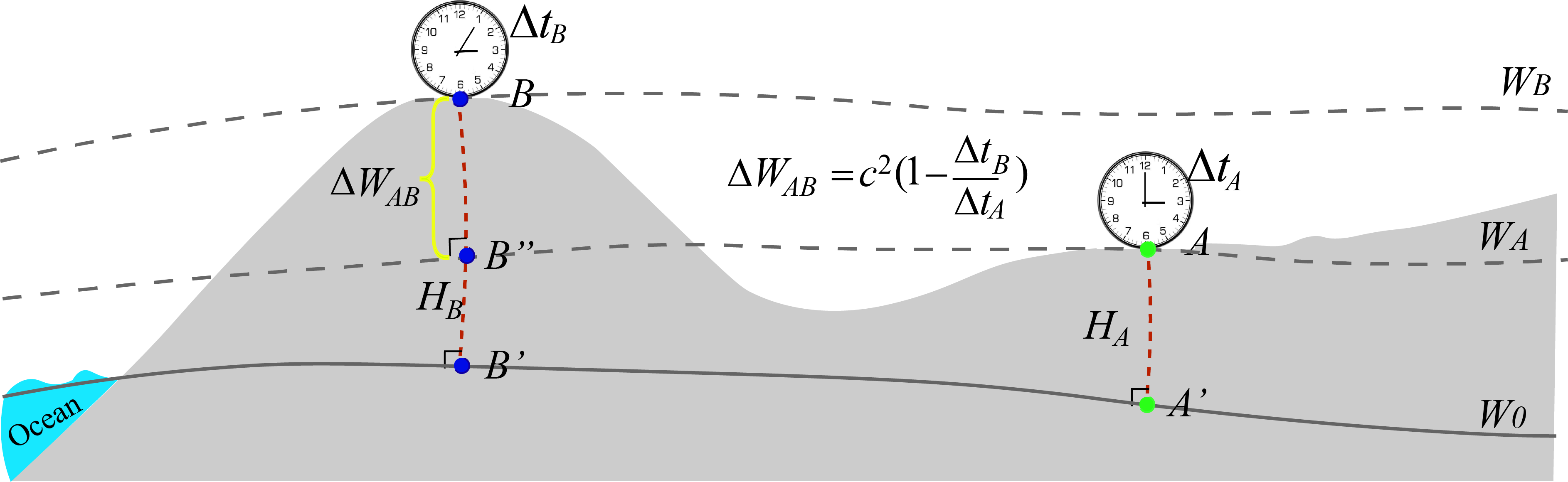}
	\caption{Schematic representation of the clock measurement.
		$W_{0}$ is geopotential on the geoid (solid curve), $\Delta t_{i}$, $W_{i}$ and $H_{i}$ denote proper time duration, geopotential, and orthometric height (OH) at position $i$ ($i=A,B$), respectively. 
		The OH $H_{i}$  (dashed red curve) is the distance between $i^{\prime}$ and $i$ along the curved plumb line.
    	$A^{\prime}$($B^{\prime}$) is the intersection point between the plumb line passing through A(B) and the geoid,
    	$B^{\prime \prime}$ is the intersection point between the plumb line $B^{\prime}B$ and the equi-geopotential surface $W_{A}$. }
	\label{fig:1}
	\end{figure}
\begin{equation}\label{eq:1}
\frac{\Delta t_{_A}/\Delta t}{\Delta t_{_B}/\Delta t}=\frac{1-c^{-2}W_{A}+O(c^{-4})}{1-c^{-2}W_{B}+O(c^{-4})}
\end{equation}
where $\Delta t$ denotes a standardized time duration (measured by a standard clock at infinity or at the center of the Earth); $\Delta t_{i}/\Delta t$ denotes the elapsed time of clock $C_{i}$ ($i=A, B$); $W_{i}$ denotes geopotential at station $i$; $c=299792458$ m/s is the speed of light in vacuum.

Here, we point out that, the deformation of the Earth surface caused by the tides are neglected, due to the fact that the tidal influences are at tens of centimeters level.  Considering the stability of the hydrogen atomic clocks used in the experiments, say at $5 \times 10^{-15}$/day level, which is equivalent to tens of meters in OH, the effect caused by the tidal influences is much smaller (two orders of magnitude lower). In the future study, however, when the clock's stability reaches $10^{-18}$ level, the tidal effects should be considered carefully.

From formula \eqref{eq:1}, the clock-comparison-determined geopotential difference between stations A and B, {\footnotesize{$\Delta W_{AB}^{(T)}$}} can be expressed as follows (accurate to $1/c^{2}$):
\begin{equation}\label{eq:2}
\Delta W_{AB}^{_{(T)}} \equiv W_{B}^{_{(T)}}-W_{A}^{_{(T)}}= c^{2}(1-\frac{\Delta t_{_B}}{\Delta t_{_A}})
\end{equation}

The OH can be determined based on the following expression \citep{Molodenskii1962Methods, Heiskanen1967Physical,Jekeli2000Heights}:
\begin{equation}\label{eq:3}
H_{i}=\frac{W_{0}-W_{i}}{\bar g_{i}}
\end{equation}
where $W_{0}$ is the geopotential on the geoid, with $W_{0}=$ 62636853.4 $\pm$0.02 m$^{2}$s$^{-2}$  \citep{S2016A, S2017Vertical},
$W_{i}$ is the geopotential at location $i$, $\bar{g}_{i}$ is the 'mean gravity value' at a particular point on the segment of the plumb line between point $i$ and $i^{\prime}$.

Suppose the geopotential $W_{A}$ at station A is a priori given, by combining formulas \eqref{eq:2} and \eqref{eq:3}, the clock-comparison-determined OH at staion B, $H_{_B}^{_{(T)}}$ can be determined as follows:
\begin{center}
	\begin{equation}\label{eq:4}
	H_{_B}^{_{(T)}}=\frac{W_{0}-(W_{A}+\Delta W_{_{AB}}^{_{(T)}})}{\bar g_{_B}}=\frac{W_{0}-W_{A}}{\bar g_{_B}}+\frac{c^{2}}{\bar g_{_B}}(\frac{\Delta t_{_B}}{\Delta t_{_A}}-1)
	\end{equation}
\end{center}

Finally, by combining formulas \eqref{eq:3} $\thicksim$ \eqref{eq:4} we can compare the discrepancy between clock-comparison-determined results ($H_{_{B}}^{_{(T)}}$) with the corresponding results determined by conventional approach ($H_{_{B}}$):
\begin{equation}\label{eq:5}
\begin{split}
D &=H_{_{B}}^{_{(T)}}-H_{_{B}}\\
&=
\left(\frac{W_{0}-W_{A}}{\bar g_{_B}}+\frac{c^{2}}{\bar g_{_B}}(\frac{\Delta t_{_B}}{\Delta t_{_A}}-1)\right)-\frac{W_{0}-W_{B}}{\bar g_{_B}} \\
&=
\frac{W_{B}-W_{A}}{\bar g_{_B}}+\frac{c^{2}}{\bar g_{_B}}(\frac{\Delta t_{_B}}{\Delta t_{_A}}-1) \\
\end{split}
\end{equation}
where $W_{B}-W_{A}$  can be determined by levelling and gravimetry or directly determined by a given Earth gravity model. In this study we use EGM2008 to obtain the geopotentials $W_{A}$ and  $W_{B}$.

The determination of {\footnotesize{$\Delta W_{AB}^{(T)}$}} as well as $H_{_{B}}^{_{(T)}}$ requires a precise measurement of time comparison. Hence, one practical solution is to set clock $C_{A}$ as the standard, after a standard time elapses $\Delta T_{A}=\Delta T$, the clock $C_{B}$ elapses $\Delta T_{B}=\Delta T'$, correspondingly. Therefore, the time elapse difference between clocks $C_{A}$ and $C_{B}$, $\alpha$ can be finally determined as follows \citep{Shen2015Geopotential}:
\begin{equation}\label{eq:6}
\alpha=\frac{\Delta t_{_B}-\Delta t_{_A}}{\Delta t_{_A}}\equiv \frac{\Delta T_{B}-\Delta T_{A}}{\Delta T_{A}}=\frac{\Delta T'-\Delta T}{\Delta T}
\end{equation}

To realize a precise time comparison between two remote clocks, we need not only atomic clocks with high stability for maintaining the standard time elapse, but also a reliable time transfer technique that can precisely measure the time elapses.


\section{Common-View Satellite Time Transfer Technique}
\label{sec:3}

\subsection{CVSTT technique}

As early as in 1980, it was proposed that the CVSTT technique could be used for comparing clock offsets between two remote clocks \citep{Allan1980Accurate, Allan1985n}. This technique was adopted by the Bureau International des Poids et Mesures (BIPM) as one of the main methods for transferring  international atomic time (TAI) signals \citep{Allan1994Technical, Imae2004Impact}. Using this method, the uncertainty of comparing remote time may achieve several nanoseconds \citep{Lewandowski1991GPS,Lewandowski1999GPS,Ray2003IGS, Hongwei2010The, Yang2012Method, Rose2016Ionospheric}. The main advantages of CVSTT technique lie in that the satellite clock errors are cancelled, and various other errors, especially ionospheric and tropospheric errors, are largely reduced due to the simultaneous two-way observations \citep{Allan1980Accurate, Yang2012Method, Defraigne2011Combining}.

In this study, we use the CVSTT technique for comparing the clock offsets. The reasons are stated as follows. Firstly, the CVSTT technique is a stable and accurate method for comparing clock offsets between two remote clocks, and this technique has been adopted by the Bureau International des Poids et Mesures (BIPM) as one of the main methods for transferring international atomic time (TAI) signals. Therefore, the results of CVSTT technique are reliable and stable. Secondly, using CVSTT technique is relative cheap and convenient in practice. Though the experimental results using optical atomic clocks via fiber links are quite accurate and stable indeed, the fiber-link is limited to a relative short distance and fixed stations with fiber links, and for a long distance even intercontinental comparison, the fiber-link is extremely challengeable and costly. In addition, the Two-way satellite time and frequency transfer (TWSTFT) technique is more accurate and stable than CWSTFT technique, but it is very expensive in economy, which could be applied in the future when the condition is available. Lastly, though the carrier phase measurements in GNSS are more precise than the corresponding precise code measurements in time transfer, the absolute clock offsets cannot be determined from carrier phase measurements due to the ambiguities issues \citep{DefraigneOn}. Therefore, the code measurements are necessary for determining the absolute clock offsets between remote time and frequency standards for a long time. Hence, at present experiment condition, the CVSTT technique and precise code measurements are adopted here.
 
\begin{figure}[hbt]
	\centering
	\includegraphics[width=1\textwidth]{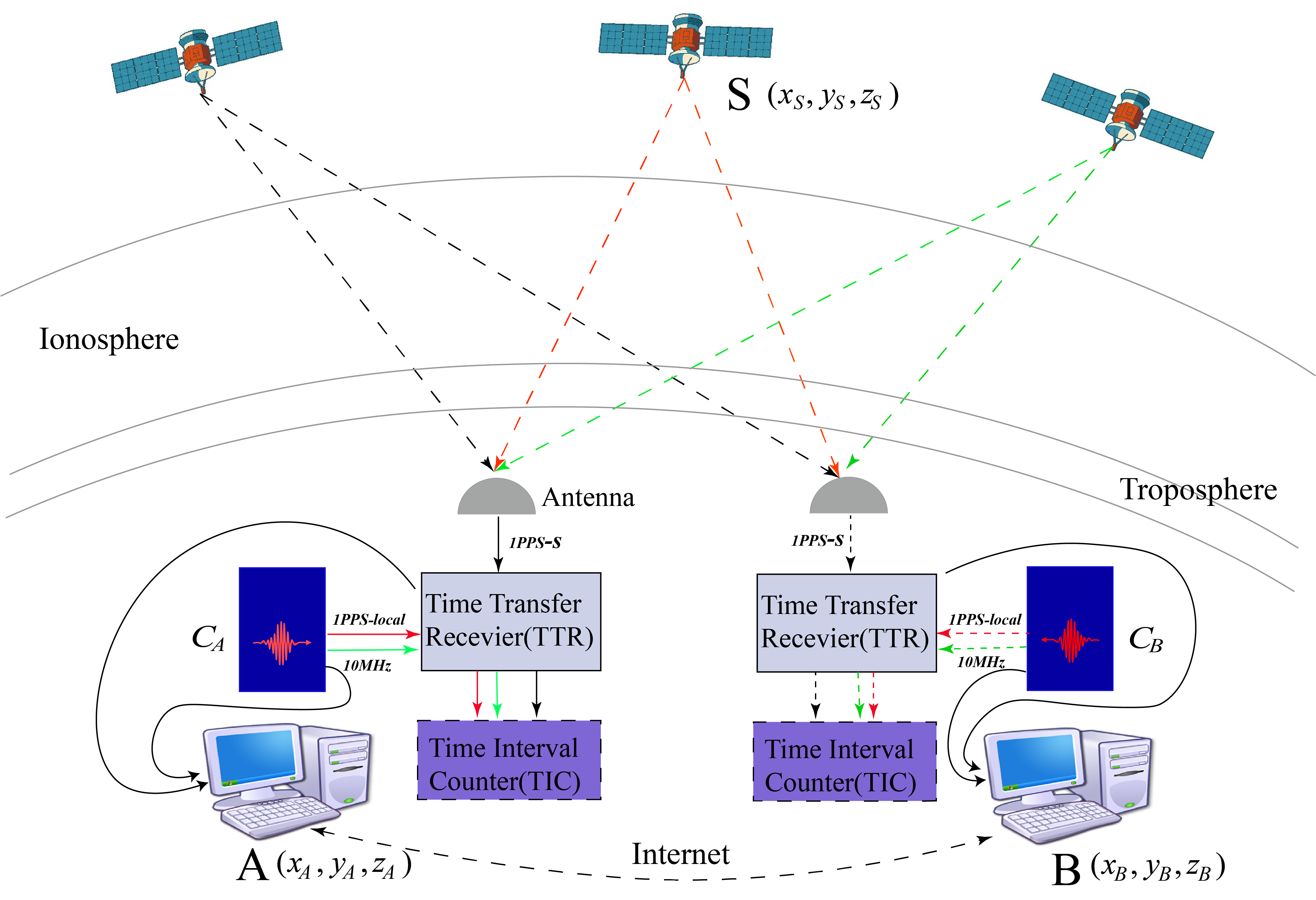}
	\caption{The schematic diagram of the CVSTT technique. The hydrogen atomic clock at one station (A or B)  generates 1 pulse per second (1PPS) signals (denoted as 1PPS-local) and 10 MHz sinusoidal signals. When the 1PPS-local signals arrive at time interval counter (TIC), the TIC  starts timing. The satellite's clock also generates 1PPS signals (denoted as 1PPS-S), simultaneously. When 1PPS-S signals arrive at TIC, the TIC stops timing. The clock offsets between the $i$-th station ($i=$A or B) and satellite $S$ can be determined by counting the number of sinusoidal waves between 1PPS-local and 1PPS-S.  After data exchange, the clock offsets between the two ground stations A and B can be finally determined by taking common-view satellite $S$ as common reference. The TIC is built-in GNSS Time Transfer Receiver (GNSS-TTR).}
	\label{fig:2}
\end{figure}

To better understand the text, here we briefly introduce the principle of the CVSTT technique.  Suppose there are two ground stations ($A$, $B$) and a common-view satellite $S$. The coordinates of the two ground stations are denoted as $(x_{i},y_{i},z_{i})$ ($i=A, B$), and the satellite coordinates are denoted as ($x_{_S}, y_{_S}, z_{_S}$). At an appointed time, the $i$-th staion's clock and satellite's clock emits 1 purse per second (1 PPS) signals, simultaneously. The 1 PPS signal of the $i$-th station (denoted as 1PPS-local signal) arrives to the GNSS time transfer receiver (GNSS-TTR) via cables, and opens the door of the time interval counter (TIC) for starting timing. At the same time, the 1PPS signal of the satellite $S$ (denoted as 1PPS-S signal) is received by the GNSS antenna of the ground station. After signals demodulation the 1PPS-S signal arrives to the GNSS-TTR via cables. And the 1PPS-S signal is used for closing the TIC's door for stopping timing.  The $i$-th staion's clock also generates the 10 MHz sinusoidal signals with extremely high accuracy. Therefore, for the $i$-th station and at a certain timepoint, by counting the number of the sinusoidal waves between the starting timepoint (the time of 1PPS-local signals arrive at TIC) and the stopping timepoint (the time of 1PPS-S signals arrive at TIC), the time interval between the two 1PPS signals can be determined as follows \citep{Allan1980Accurate, Chang1990Development, Lewandowski1991GPS}:
\begin{equation}\label{eq:7}
T_{i}=t_{i}-t_{s}+\rho_{is}/c+\delta\rho_{is}/c+t_{r_{i}}
\end{equation}
where $T_{i}$ is time interval measurement of the two 1 PPS signals at station $i$; $t_{i}$ and $t_{s}$ are the emission timepoints of 1PPS signals of the atomic clocks from $i$-th station and satellite $S$, respectively; $\delta\rho_{is}$ is the error source (e.g. ionospheric delay, tropospheric delay, and Sagnac effect); $t_{r_{i}}$ is the receiver delay which is usually calibrated by the manufacturing company using simulated signals \citep{Lewandowski1991GPS}; and $\rho_{is}$ is the geometric distance between $i$-th station and satellite $S$, expressed as:
\begin{equation}\label{eq:8}
\rho_{is}=\sqrt{(x_{_S}-x_{i})^{2}+(y_{_S}-y_{i})^{2}+(z_{_S}-z_{i})^{2}}
\end{equation}
By applying equation \eqref{eq:7}, the clock offsets between the $i$-th station and satellite S can be expressed as follows:
\begin{equation}\label{eq:9}
\Delta t_{si}=t_{i}-t_{s}=T_{i}-\rho_{is}/c-\delta\rho_{is}/c-t_{r_{i}}
\end{equation}
Finally, the clock offsets between two ground stations A and B can be determined by taking the common-view satellite's clock  as common reference:
\begin{equation}\label{eq:10}
\begin{split}
\Delta t_{_{AB}}&=t_{_{B}}-t_{_{A}}\\
&=
(t_{_S} + \Delta t_{_{SB}})- (t_{_S} + \Delta t_{_{SA}})\\
&=
(T_{_B}-\frac{\rho_{_{BS}}}{c}-\frac{\delta\rho_{_{BS}}}{c}-t_{r_{_B}})-(T_{_A}-\frac{\rho_{_{AS}}}{c}-\frac{\delta\rho_{_{AS}}}{c}-t_{r_{_A}}) \\
&= (T_{_B}-T_{_A})-\frac{(\rho_{_{BS}}-\rho_{_{AS}})}{c}-\frac{(\delta\rho_{_{BS}}-\delta\rho_{_{AS}})}{c}-(t_{r_{_B}}-t_{r_{_A}}) \\
\end{split}
\end{equation}

We note that the cable delays from station's atomic clock to TTR are measured beforehand with the value of 50.2 ns, as well as the cable delays from GNSS antenna to TTR with the value of 204.5 ns. And the cable delays are input into the TTR as a parameter. In fact, the cable delays are not significant in our study due  to the fact  that we compare the time elapse between two remote clocks. 

\subsection{Error sources}
\label{sec:3.2}
\subsubsection{Satellite position Error}
\label{sec:3.2.1}
In this study, the transportation experiments is conducted at Beijing 203 Institute Laboratory (BIL) and Luojiashan Time-Frequency Station (LTS). In the sequel, we analyze various  error sources.

The accuracy of the satellite positions depends on the ephemeris.  The broadcast ephemeris is adopted in conventional CVSTT technique. Suppose the satellite position error is ($\delta x_{s}$, $ \delta y_{s}$, $ \delta z_{s}$), when ignoring the position error of ground stations, the relation between satellite position error and time transfer error in CVSTT technique can be determined by using the first-order difference \citep{Imae2004Impact, Li2008ERROR, Hongwei2010The}:
\begin{equation}\label{eq:11}
\Delta \tau_{_{AB}}=(l_{_{AS}}-l_{_{BS}})\frac{\delta x_{s}}{c}+(m_{_{AS}}-m_{_{BS}})\frac{\delta y_{s}}{c}+(n_{_{AS}}-n_{_{BS}})\frac{
	\delta z_{s}}{c}
\end{equation}
where
\begin{equation}\label{eq:12}
\begin{split}
\begin{aligned}
&l_{is}=\frac{x_{s}-x_{i}}{\rho_{is}}     \\
&m_{is}=\frac{y_{s}-y_{i}}{\rho_{is}}   \\
&n_{is}=\frac{z_{s}-z_{i}}{\rho_{is}}    \\
\end{aligned}
\end{split}
\end{equation}

Based on formulas (\ref{eq:11}) and (\ref{eq:12}) and applying  the error propagation law, the following expression can be obtained:

\begin{equation}\label{eq:13}
m^{2}(\Delta \tau_{_{AB}})=(\frac{l_{_{AS}}-l_{_{BS}}}{c})^{2}m^{2}(\delta x_{s}) +	(\frac{m_{_{AS}}-m_{_{BS}}}{c})^{2}m^{2}(\delta y_{s})+(\frac{n_{_{AS}}-n_{_{BS}}}{c})^{2}m^{2}(\delta z_{s})
\end{equation}

For instance, when accuracy of broadcast ephemeris in individual axes is 2 m \citep{Gao2004Use, Liu2017Error}, the accuracy of time transfer error caused by satellite position error did not exceed 0.16 ns in the experiments. The influences of the ground stations' positions errors on the time transfer in CVSTT technique can be evaluated, similarly. Here the ground stations' coordinates are determined based on the precise point positioning (PPP) method in advance, with accuracy level of better than 3 cm \citep{DawidowiczCoordinate}, and the corresponding time transfer error caused by the coordinates errors will not exceed 0.15 ns (see Table \ref{tab:1}).

\subsubsection{Ionospheric effect}
\label{sec:3.2.2}
The ionospheric effect on GNSS signal means a signal delay which could achieve several tens of nanoseconds, due to the electron content of the atmospheric layer. And the effect becomes more dramatic during an ionospheric storm.  The zenith group delays are about 1 $\thicksim$ 30 m, 0$\thicksim$2 cm, and 0$\thicksim$2 mm for the first-, second- and third-order ionospheric effects, respectively \citep{Bassiri1993Higher, Marques2011RINEX, Zhang2013}. For the GNSS time transfer receivers, measurements on two frequencies ($f_{1}$ and $f_{2}$) are often available. Therefore, an ionosphere-free observation $P_{3}$ can be constructed so that the  first-order term of ionospheric  effect can be removed completely, due to the fact that the ionospheric effect on a signal depends on its frequency. The $P_{3}$ is constructed as follows \citep{Petit2009Use, Peng2009GPS}:
\begin{equation}\label{eq:14}
P_{3}=\frac{1}{f_{1}^2-f_{2}^2}(f_{1}^2P_{1}-f_{2}^2P_{2})
\end{equation}
where $P_{1}$ and $P_{2}$ are the precise-code observations with frequencies of $f_{1}=1575.42$ MHz and $f_{2}=1227.60$MHz, respectively. 

However, formula (\ref{eq:14}) only removing the first-order ionospheric effect. There are second- and third-order ionospheric effects (also denoted as higher-order ionospheric effects). The residual higher-order ionospheric effects, $I_{r}$, can be expressed as follows \citep{Morton2009Assessment}:
\begin{equation}\label{eq:15}
\begin{split}
&I_{r}=\frac{q}{f_{1}f_{2}(f_{1}+f_{2})}+\frac{t}{f_{1}^{2}f_{2}^{2}} \\
&q=2.2566 \times 10^{12}\int N_{e}B_{0}cos\theta_{_B}\,ds \\
&t=2437\int N_{e}^{2}\,ds+ 4.74 \times 10^{22} \int N_{e}B_{0}^{2}(1+cos^{2}\theta_{_B})\,ds \\
\end{split}
\end{equation}
where $N_{e}$ is the number of electrons in a unit volume($/m^{3}$), $B_{0}$ is the magnitude of the plasma magnetic field(T), $\theta_{_B}$ is the angle between the wave propagation direction and the local magnetic field direction, and $s$ is the integral path. More details can be seen in \cite{Morton2009Assessment} and \cite{Zhang2013}.

By using CVSTT technique, the residual errors of ionospheric effects between ground stations A and B, $I_{r_{(AB)}}$, can be determined as follows:
\begin{equation}\label{eq:16}
I_{r_{(AB)}}=\frac{q_{_A}-q_{_B}}{f_{1}f_{2}(f_{1}+f_{2})}+\frac{t_{_A}-t_{_B}}{f_{1}^{2}f_{2}^{2}}
\end{equation}

By using the constructed ionosphere-free observation $P_{3}$, we remove the first-order ionospheric effect, which contributes to more than 99\% of the entire ionospheric effects \citep{Defraigne2010Combining, Harmegnies2013Combining, Huang2016BeiDou}, and the residual errors of higher-order terms are about 2$\thicksim$4 cm \citep{Bassiri1993Higher, Morton2009Assessment, Deng2015Analysis}. 

On the other hand, the code noise should be taken into consideration carefully. Suppose code noise of $P_{1}$ and $P_{2}$ are $m_{1}$ and $m_{2}$, respectively. By applying the error propagation law, the code noise of ionosphere-free observation, $m_{3}$, can be expressed as follows:
\begin{equation}\label{eq:17}
m_{3}^{2}=(\frac{f_{1}^2}{f_{1}^2-f_{2}^2})^{2}m_{1}^{2}+(\frac{f_{2}^2}{f_{1}^2-f_{2}^2})^{2}m_{2}^{2}
\end{equation}

Here, suppose the code noise $m_{1}\approx m_{2}$, and take the values of $f_{1}=1575.42$ MHz and $f_{2}=1227.60$ MHz into formula (\ref{eq:17}), the relations of $m_{3}\approx 2.98m_{1}$ can be obtained. The accuracy of P code measurement is about 0.29m, which is equivalent to 0.97 ns in time. Therefore, the accuracy 2.89 ns of  the $P_{3}$ code measurement is determined, and the corresponding time transfer accuracy of the time interval measurement in CVSTT technique, $T_{B}-T_{A}$, will not exceed 4.09 ns. Combining expressions  (\ref{eq:16}) and (\ref{eq:17}), after implementation of strategy of ionosphere-free combination, the residual ionospheric errors $I_{r_{(AB)}}$ in CVSTT technique will not exceed 0.8 ns (see Table \ref{tab:1}).

\subsubsection{Tropospheric effect}
\label{sec:3.2.3}
The troposphere is the lower layer of atmosphere that extends from ground to base of the ionosphere. The signal transmission delay caused by troposphere is about 2$\thicksim$20 m from zenith to horizontal direction \citep{Xu2007GPS, Chen2012An}. The tropospheric delay depends on temperature, pressure, humidity, and as well as location of the GPS antenna \citep{quteprints23998}. The total tropospheric delay can be divided into hydrostatic part and wet part, which can be expressed as follows \citep{Boehm2004Vienna, Kouba2008Implementation}:
\begin{equation}
\label{eq:18}
\Delta L=\Delta L_{h}^{z}\cdot mf_{h}(E, a_{h}, b_{h}, c_{h})+\Delta L_{w}^{z}\cdot mf_{w}(E, a_{w}, b_{w}, c_{w})
\end{equation}
where $\Delta L$ is the total tropospheric delay, $\Delta L_{h}^{z}$ and $\Delta L_{w}^{z}$ are the hydrostatic part and wet part in zenith delays, respectively. $mf_{h}$ and $mf_{w}$ are the corresponding mapping functions (MF), respectively. MF is a function of $E$, $a_{h(w)}$, $b_{h(w)}$, and $c_{h(w)}$, E is the elevation angle in radian. $a_{h(w)}$, $b_{h(w)}$, and $c_{h(w)}$ are MF coefficients.

\cite{Boehm2004Vienna} introduced a rigorous approach for utilizing the Numerical Weather Models (MWM) for MF determinations. The first and most significant MF coefficients, $a_{h}$ and $a_{w}$, are fitted with the NWM from the European Centre for Medium-Range Weather Forecasts (ECMWF). The coefficient $b_{h}=0.002905$, and coefficient $c_{h}$ is expressed as follows:
\begin{equation}\label{eq:19}
c_{h}=c_{0}+\left[\left(cos\left(\frac{doy-28}{365} \cdot 2\pi +\Psi\right)+1\right)\cdot \frac{c_{11}}{2}+c_{10}\right]\cdot (1-cos \varphi)
\end{equation}
where $c_{0}=0.062$ is the day of the year; $\varphi$ is the latitude; $\Psi$, $c_{10}$, and $c_{11}$ specifies the Northern or Southern Hemisphere, respectively. For the wet part, the coefficients $b_{w}$ and $c_{w}$ are constants with $b_{w}=0.00146$ and $c_{w}=0.04391$ \citep{Boehm2006}.

Then, the gridded VMF1 data is generated from the ECMWF NWM using four global grid files($2.0^{\circ} \times 2.5^{\circ}$) of $a_{h}$, $a_{w}$, $\Delta L_{h}^{z}$ and $\Delta L_{w}^{z}$. Using VMF1, one can determine the tropospheric delay at any location after 1994 \citep{Kouba2008Implementation}. The tar-compressed files of all VMF1 epoch grid files are available at the VMF1 website (\url {http://ggosatm.hg.tuwien.ac.at/DELAY/GRID/VMFG/}). More details are available in \cite{Boehm2004Vienna} and \cite{Kouba2008Implementation}.
\begin{figure}[hbt]
	\centering
	\includegraphics[width=1\textwidth]{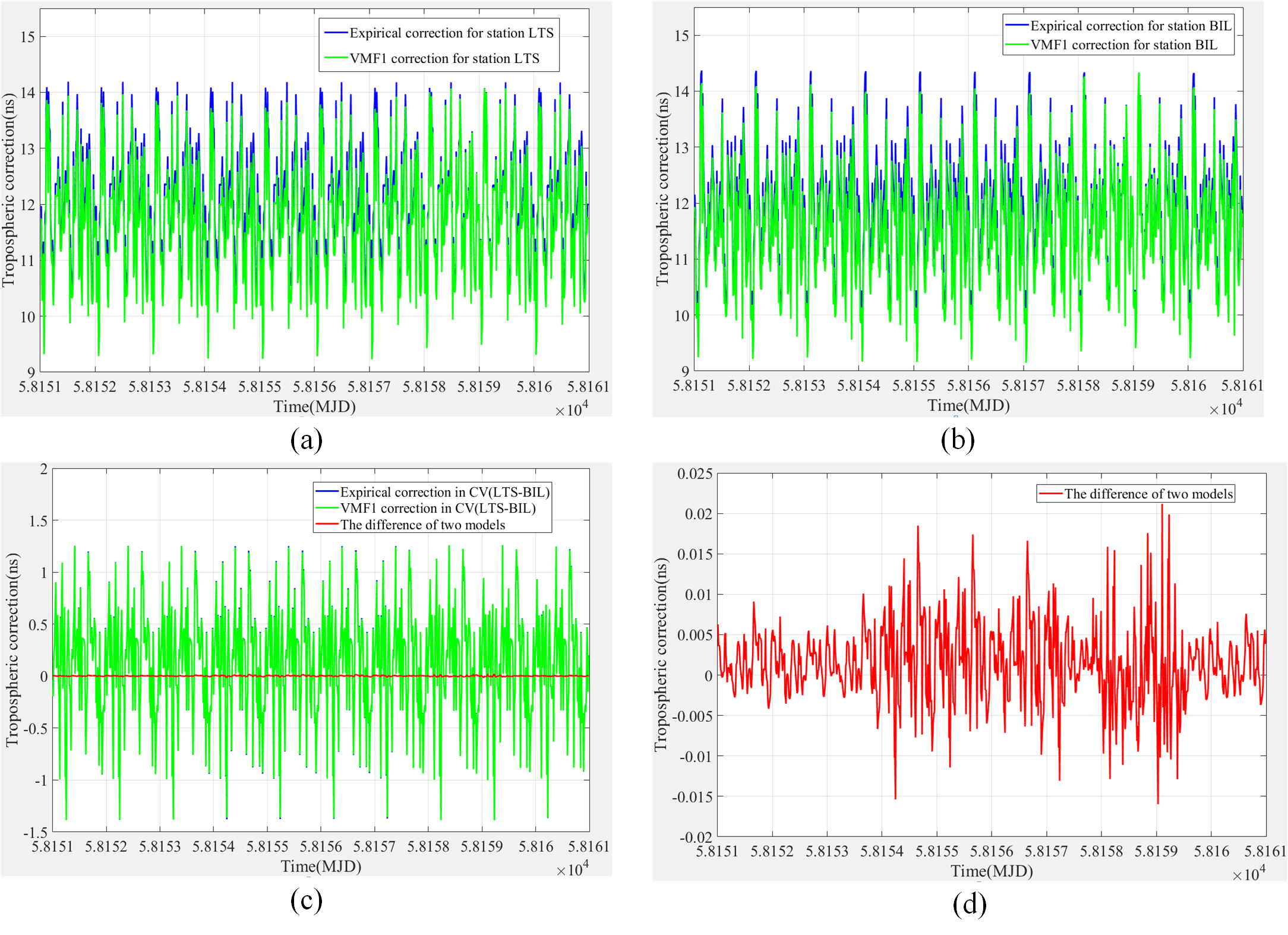}
	\caption{The tropospheric corrections between stations BIL and LTS, which last from MJD 58151 to MJD 58160. The tropospheric corrections for stations LTS and BIL are shown in (a) and (b), respectively. (c) is tropospheric corrections for the two stations via CVSTT technique. And (d) is the difference between empirical model and VMFI model. It should be noted that the blue curve denotes tropospheric corrections by the empirical model, the green curve denotes tropospheric corrections by the VMF1 model, and the red curve denotes the difference of the two models.
	}
	\label{fig:3}
\end{figure}

To achieve near real-time data processing and improve efficiency, a simple empirical model is employed, which is expressed as follows \citep{Parkingson1996}:
\begin{equation}\label{eq:20}
\Delta L_{{trop}}=2.47/(sin E+0.0121)
\end{equation}

 In order to verify the accuracy of the empirical model, we take the two stations (BIL and LTS) from MJD 58151 to 58160 into calculation. We use VMF1 as a comparison, and compare the correction magnitude of the two models in CVSTT technique. The results are shown in Fig. \ref{fig:3}, the elevation set-off in our experiment is $20^{\circ}$. From Fig. \ref{fig:3}, we conclude that the correction magnitude is tens of nanoseconds for individual stations, and about 2 ns for the two stations via CVSTT technique. And the difference between empirical model and VMF1 is quite small for CVSTT technique (less than 0.025 ns). Therefore, the empirical model is precise enough in our study. In our experiments, after using empirical model for tropospheric correction, the accuracy of the correction achieves 0.52 ns (see Table \ref{tab:1}).

\subsubsection{Sagnac effect}
\label{sec: 3.2.4}

The Sagnac effect is caused by the rotation of the Earth. During the signal propagation from satellite $S$ to station $i$, the Sagnac effect (correction)  is expressed as \citep{Ashby2004The, Tseng2011Sagnac, Yang2012Method}:
\begin{equation}\label{eq:21}
S{_i}=- \omega_{e}\frac{x_{_S}y_{i}-x_{i}y_{_S}}{c^{2}}
\end{equation}
where $\omega_{e}$ is the Earth rotation rate, with a relative uncertainty of $\delta \omega_{e}=1.4 \times 10^{-8}$ \citep{Groten2010}. 

Taking station A and station B into considertion, the corrections of signal delay of the total Sagnac effect via CVSTT technique can be expressed as follows:
\begin{equation}\label{eq:22}
\begin{split}
\Delta S{_{_{AB}}}&=S_{_B}- S_{_A} \\
&=
[y_{_S}(x_{_B}-x_{_A})-x_{_S}(y_{_B}-y_{_ A})] \omega_{e}/c^{2} \\
\end{split}
\end{equation}

And the accuracy of Sagnac corrections via CVSTT technique can be determined through error propagation law, which is expressed as:
\begin{equation}\label{eq:23}
\begin{split}
m^{2}_{_{Sac}}&=[(x_{_B}-x_{_A})^{2}m^{2}(\delta y_{_S})+(y_{_B}-y_{_A})^{2}m^{2}(\delta x_{_S}) \\
&+
y_{_S}^{2}m^{2}(\delta x_{_A})+y_{_S}^{2}m^{2}(\delta x_{_B})_+x_{_S}^{2}m^{2}(\delta y_{_A})+x_{_S}^{2}m^{2}(\delta y_{_B})](\omega_{e}/c^{2})^{2} \\
&+
[y_{_S}(x_{_B}-x_{_A})-x_{_S}(y_{_B}-y_{_A})] m^{2}(\delta \omega_{e})/c^{2}\ \\
\end{split}
\end{equation}

Taking the coordinates of BIL and LTS into consideration, results show that $m_{_{Sac}}$ will not exceed 0.003 ns.

\begin{table*}
	\caption{The accuracy of CVSTT technique due to different error sources and the corresponding accuracy of OH determiation in the transportation experiments.}
	\label{tab:1}
	\begin{center}
		\setlength{\tabcolsep}{9mm}{
			\begin{tabular}{lllllll}
				\hline\noalign{\smallskip}
				Error type  & Sources errors  & CVSTT accuracy(ns) & OH accuracy(m) \\
				\noalign{\smallskip}\hline\noalign{\smallskip}
				code noise                        & 0.97 ns & 4.09 &43.42\\
				Broadcast ephemeris         & 2 m & 0.16 &1.70\\
				Ground station                 &3 cm & 0.15 &1.60\\
				Ionosphere                       & - & 0.80 &8.50\\
				Troposphere                    & - & 0.52 &5.52\\
				Sagnac                            & $1.4\times10^{-18}$   & 0.003 &0.04\\
				\noalign{\smallskip}\hline\noalign{\smallskip}
				Total                                &  -   &4.21         &44.70\\
				\noalign{\smallskip}\hline
		\end{tabular}}
	\end{center}
\end{table*}


\section{Data processsing method}
\label{sec: 4}
\subsection{EEMD technique}
\label{sec:  4.1}
Previous studies \citep{Shen2014Observation} demonstrates that the ensemble empirical mode decomposition (EEMD) \citep{Zhaohua2009ENSEMBLE} is an effective technique for isolating target signals from environmental noises. The EEMD technique is developed from the empirical mode decomposition (EMD) \citep{Huang1998Huang} (\url {https://royalsocietypublishing.org/doi/pdf/10.1098/rspa.1998.0193}). The signals series are decomposed into a series of Intrinsic Mode Functions (IMFs) and a residual trend $r$  after EEMD decomposition. These IMFs series that are sifted stage by stage reflect local characteristics of the signals, while the residual trend $r$ series reflects slow change of the signals \citep{Zhu2013Harmonic}. The specific steps of EEMD are stated as follows \citep{Zhaohua2009ENSEMBLE} (\url {https://www.worldscientific.com/doi/abs/10.1142/S1793536909000047}):

1) The overall signal series $X(t)$ is determined after adding Gaussian white noise $\omega(t)$ into original signal series $x(t)$:
\begin{equation}\label{eq:24}
X(t)=x(t)+\omega(t)
\end{equation}

2) The EMD method is applied to implement decomposition for $X(t)$ subsequently, these IMFs are determined as:
\begin{equation}\label{eq:25}
X(t)=\sum_{j=1}^nc_{j}+r_{n}
\end{equation}
where $c_{j}$ denotes IMF$_{j}$ and $r_{n}$ denotes the residual trend.

3) Adding different Gaussian white noise $\omega_{i}(t)$ into $x(t)$ and repeating step 1) and step 2), a series of $c_{ij}$ and $r_{in}$ are determined:
\begin{equation}\label{eq:26}
X_{i}(t)=\sum_{j=1}^nc_{ij}+r_{in}
\end{equation}

4) According to zero mean principle of Gaussian white noise, the interference cuased by Gaussian white noise can be cancelled after averaging.Therefore, the IMF series $c_{n}(t)$ can be determined as:
\begin{equation}\label{eq:27}
c_{n}(t)=\frac{1}{N}\sum_{i=1}^nc_{i,n}(t)
\end{equation}

Finally, the original signal series $x(t)$ is decomposed into a series of IMFs series $c_{n}(t)$ and a residual trend series $r(t)$ by EEMD technique:
\begin{equation}\label{eq:28}
x(t)=\sum_{n=1}^mc_{n}(t)+r(t)
\end{equation}

After EEMD decomposition of the signal series $x(t)$, the index of the orthogonality (IO) is calculated to check the completeness of the decomposition, which is defined as follows \citep{Huang1998Huang}:
\begin{equation}\label{eq:29}
IO=\sum_{t=0}^{T}(\sum_{j=1}^{n+1}\sum_{k=1}^{n+1}c_{j}(t)c_{k}(t)/X^{2}(t))
\end{equation}
where $X^{2}(t)$ is defined as:
\begin{equation}\label{eq:30}
X^{2}(t)=\sum_{j=1}^{n+1}c_{i}^{2}(t)+2\sum_{j=1}^{n+1}\sum_{k=1}^{n+1}c_{j}(t)c_{k}(t)
\end{equation}

If the decomposition is completely orthogonal, the cross terms in the right-hand side of  expression \eqref{eq:29} should be zero, namely IO$=0$; and for the worst case IO$=1$.

\subsection{Simulation experiments}
\label{sec: 4.2}
Here, using a simulation experiments we explain the advantages of the EEMD technique. Suppose we have a synthetic series $S_{base}(t)$ that consists of 3 periodic signals series,$S_{i}(t)=A_{i}\cdot \sin(2\pi f_{i}t) \cdot \exp(-10^{-7}t)$(units:ns), where $A_{1}=10$, $A_{2}=A_{3}=5 $, $f_{1}=2/86400$ Hz, $f_{2}=1/86400$ Hz, $f_{3}=0.5/86400$ Hz, respectively, and a linear signal series, $S_{4}(t)= 3 \cdot 10^{-4}t-10^{-7}$, with a data length of 10 days and sampling interval 960 seconds. The $S_{base}(t)$ can be expressed as:
\begin{equation}\label{eq:31}
\begin{split}
S_{base}(t)&=S_{1}(t)+S_{2}(t)+S_{3}(t)+S_{4}(t) \\
&=
10\cdot \sin(2\pi\cdot\dfrac{2}{86400}\cdot t) \cdot \exp(-10^{-7}t)+5\cdot \sin(2\pi\cdot\dfrac{1}{86400}\cdot t) \cdot \exp(-10^{-7}t) \\
&+5\cdot \sin(2\pi\cdot\dfrac{0.5}{86400}\cdot t) \cdot \exp(-10^{-7}t) + (3 \cdot 10^{-4}\cdot t-10^{-7}) \\
\end{split}
\end{equation}
\begin{figure}[hbt]
	\centering
	\includegraphics[width=1\textwidth]{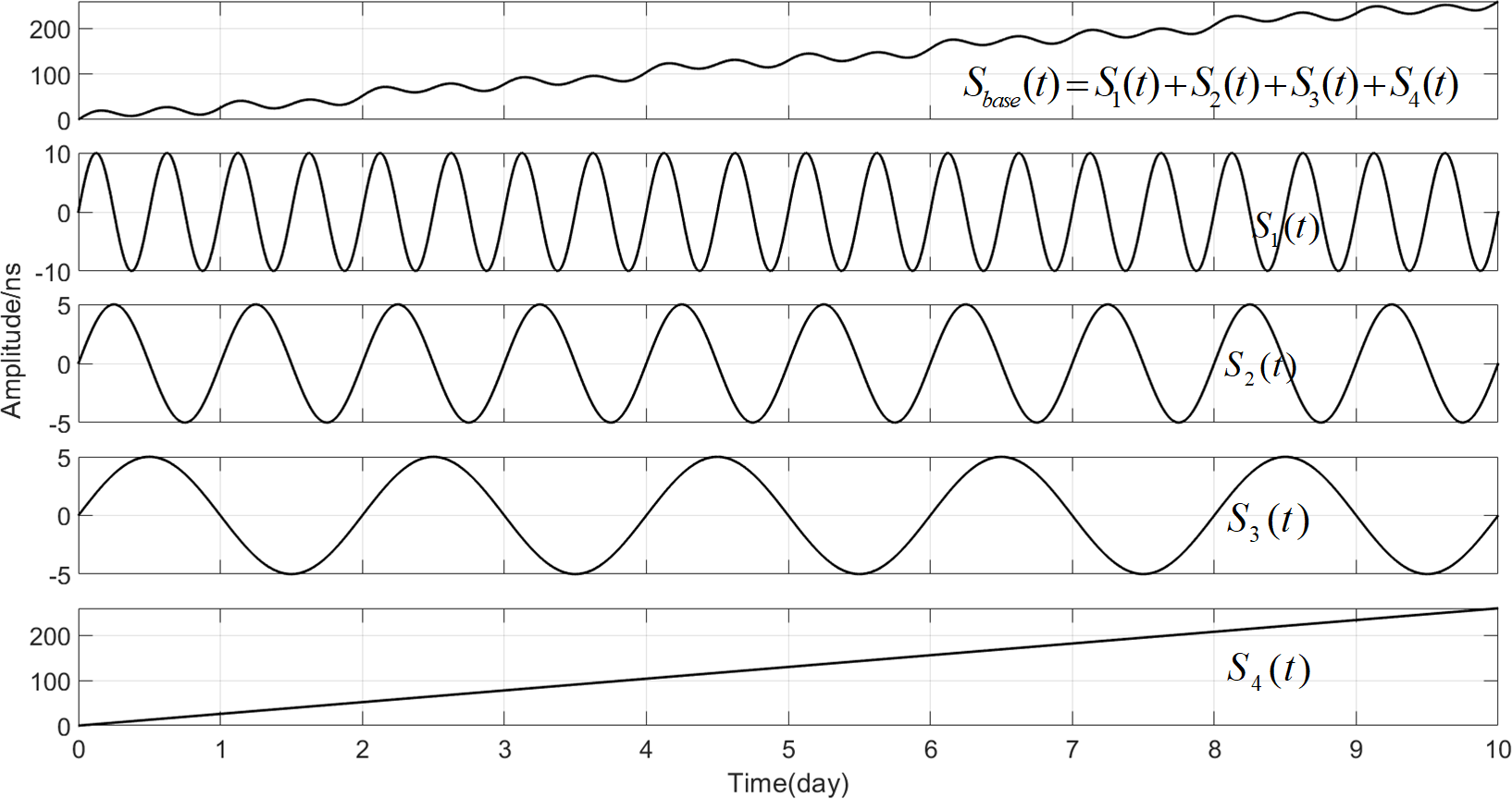}
	\caption{The waveforms of the synthetic series $S_{base}(t)$ (top slot) and different components $S_{i}(t)$ ($i$=1,2,3,4) in the subsequent slots, respectively.}
	\label{fig:4}
\end{figure}
\begin{figure}[hbt]
	\centering
	\includegraphics[width=1\textwidth]{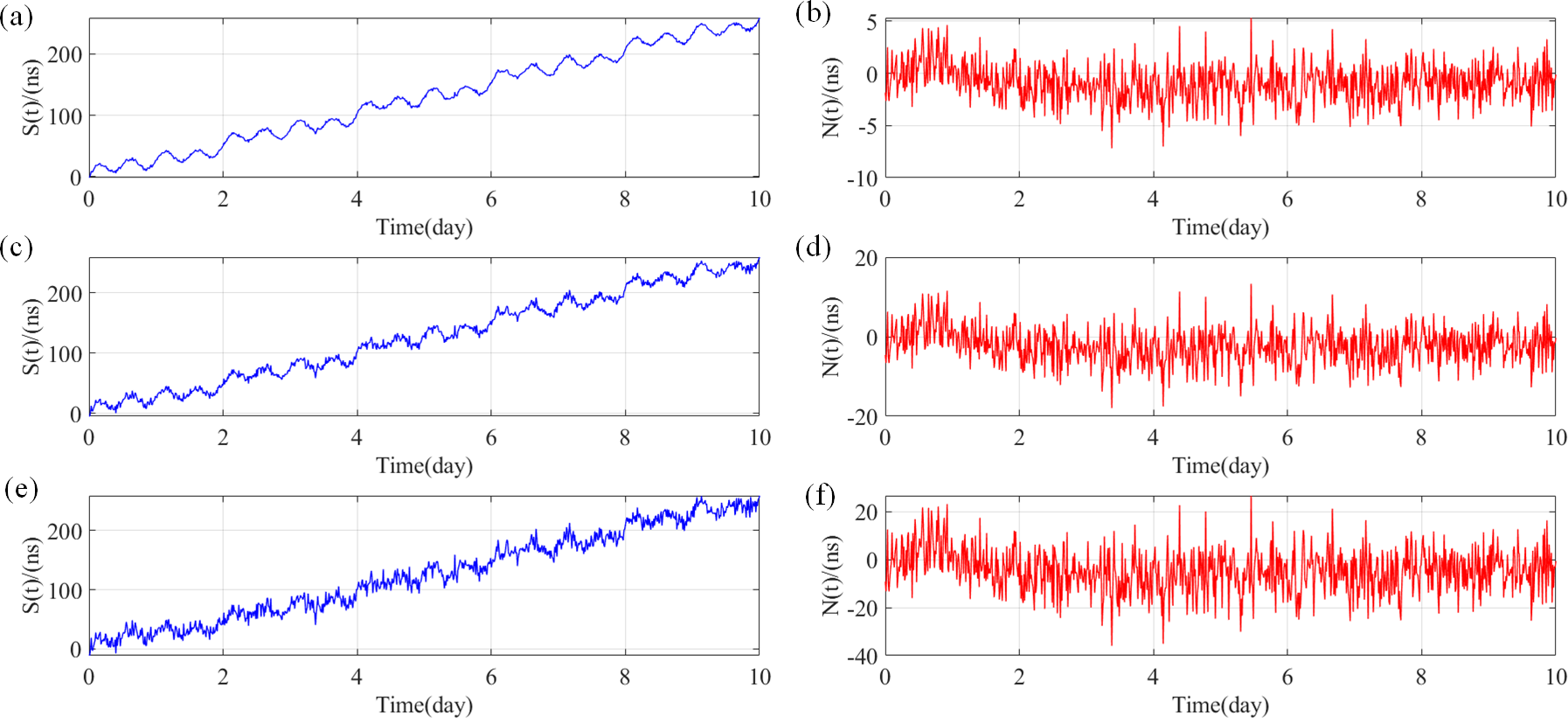}
	\caption{The waveforms of the constructed signal series $S(t)$ and different magnitudes of the noise series $N(t)$. (a), (c) and (e) are the constructed signal series $S(t)$ which are the sum of $S_{base}(t)$ and corresponding $N(t)$ with magnitude of 2$\%$, 5$\%$ and 10$\%$ of $S_{base}(t)$, respectively; (b), (d) and (f) are the noise series $N(t)$ with magnitudes of  2$\%$, 5$\%$ and 10$\%$ of $S_{base}(t)$, respectively.}
	\label{fig:5}
\end{figure}

 The results of the constructed series are shown in  Fig. \ref{fig:4}. We added noise signals $N(t)$ into $S_{base}(t)$ with 2$\%$ (Case 1), 5$\%$ (Case 2) and 10$\%$ (Case 3) of the standard deviation (STD) of $S_{base}(t)$, and then construct a signal series $S(t)$ which contains noise. The $S(t)$ can be expressed as:
\begin{equation}\label{eq:32}
S(t)=S_{base}(t)+N(t)
\end{equation}
where the $N(t)$ is added noise, which consists of five types of noises, expressed as \citep{Simulation2011,Different2012,Probability2015}:
\begin{equation}\label{eq:33}
N(t)=N_{W-PM}(t)+N_{F-PM}(t)+N_{W-FM}(t)+N_{F-FM}(t)+N_{RW-FM}(t)
\end{equation}
where $N_{W-PM}(t)$ is the white noise phase modulation (W-PM), $N_{F-PM}(t)$ is the flicker noise phase modulation (F-PM), $N_{W-FM}(t)$ is the white noise frequency modulation (W-FM), $N_{F-FM}(t)$ is the flicker noise frequency modulation (F-FM), $N_{RW-FM}(t)$ is the random walk noise frequency modulation (RW-FM).

The constructed signal series $S(t)$ with different noise magnitudes are shown in Fig. \ref{fig:5}. For our present purpose, the linear signal series $S_{4}(t)$ is the target signal which need to be extracted from the synthetic signals series $S(t)$.

\begin{figure}[hbt]
	\centering
	\includegraphics[width=1\textwidth]{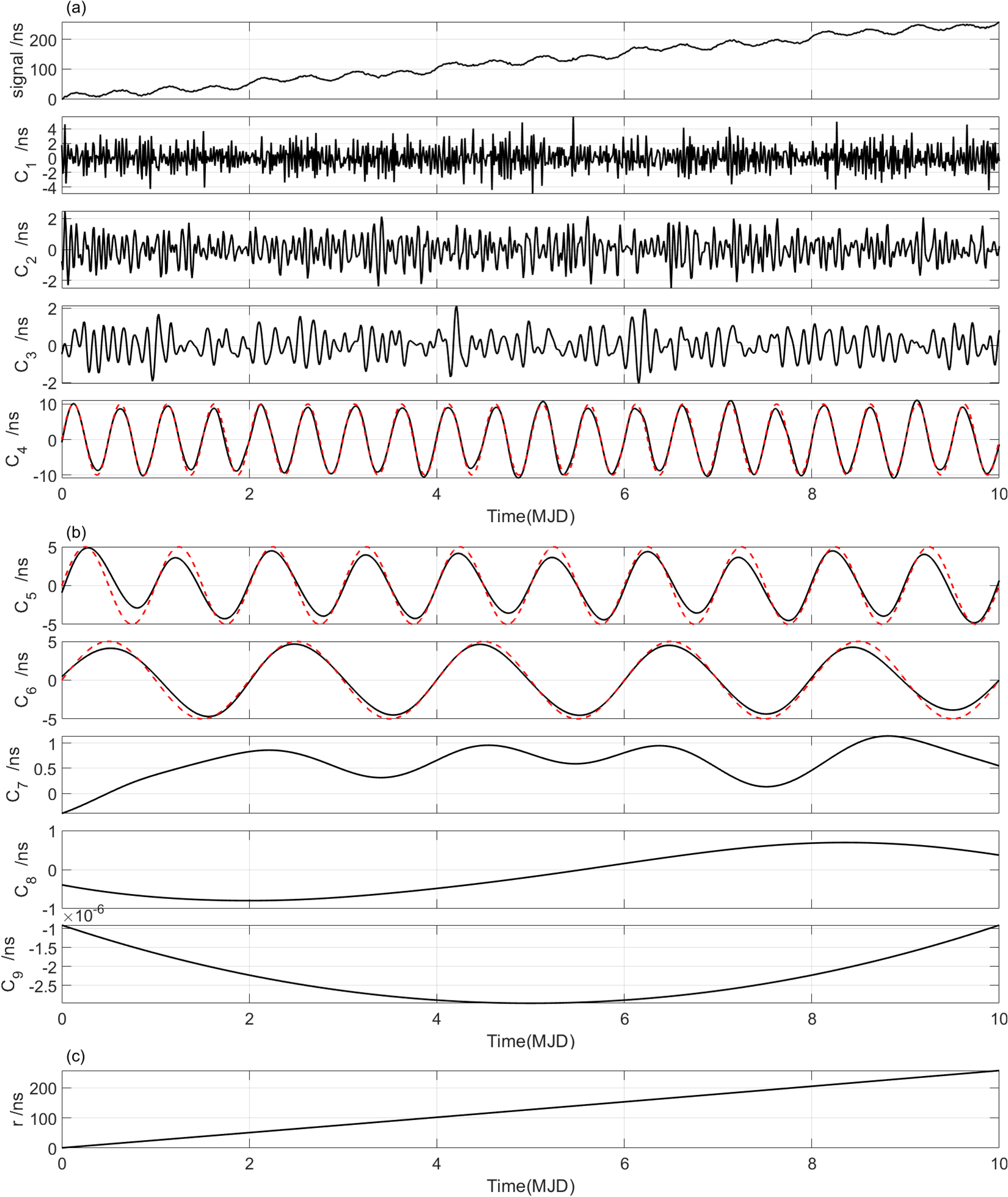}
	\caption{The results of IMFs from the constructed signal series $S(t)$ in Case 1 (with a 2$\%$ of the magnitude of the noise). (a) The constructed signal series $S(t)$ and the components IMF1 $\thicksim$ IMF4 (black curves), (b) the components IMF5 $\thicksim$ IMF9 (black curves), (c) the residual trend r. Dotted red curves are original signals.}
	\label{fig:6}
\end{figure}

\begin{figure}[hbt]
	\centering
	\includegraphics[width=1\textwidth]{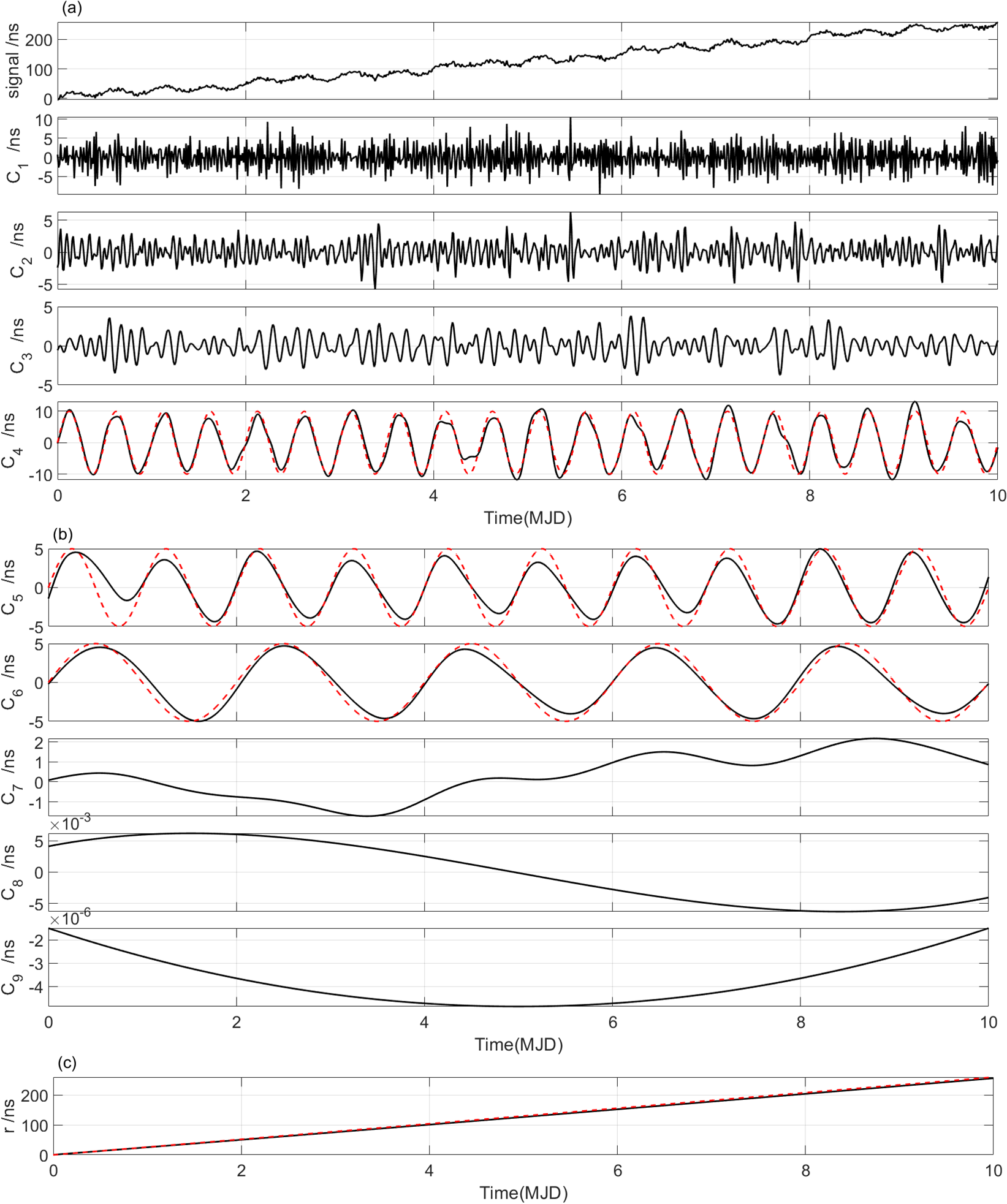}
	\caption{The resulting IMFs from the constructed signal series $S(t)$ in Case 2 (with a 5$\%$ of the magnitude of the noise). (a) The constructed signal series $S(t)$ and the components IMF1 $\thicksim$ IMF4 (black curves), (b) the components IMF5 $\thicksim$ IMF9 (black curves), (c) the residual trend r. Dotted red curves are original signals.}
	\label{fig:7}
\end{figure}

\begin{figure}[hbt]
	\centering
	\includegraphics[width=1\textwidth]{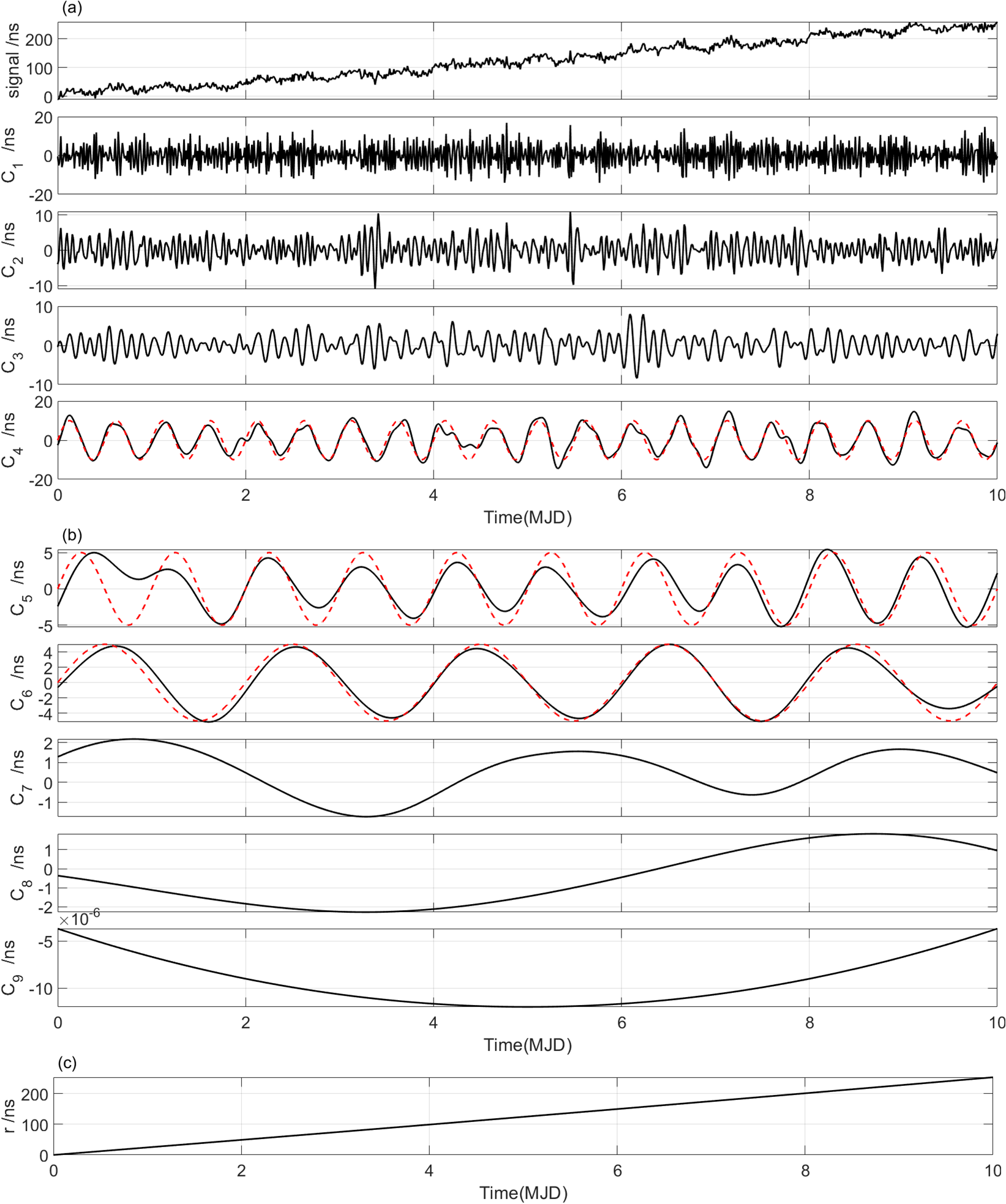}
	\caption{The resulting IMFs from the constructed signal series $S(t)$ in Case 3 (with a 10$\%$ of the magnitude of the noise). (a) The constructed signal series $S(t)$ and the components IMF1 $\thicksim$ IMF4  (black curves), (b) the components IMF5 $\thicksim$ IMF9 (black curves), (c) the residual trend r. Dotted red curves are original signals.}
	\label{fig:8}
\end{figure}

The EEMD technique is applied to identify the periodic signals $S_{1}(t)$, $S_{2}(t)$ and $S_{3}(t)$ from $S(t)$. After EEMD decomposition, we use the index IO to check completeness of the decomposition. In our simulation experiments, the values of IO are 0.0015 (Case 1), 0.0017 (Case 2) and 0.0017 (Case 3), respectively, suggesting that signal series $S(t)$ are effectively decomposed in three different cases. The decomposed IMFs are shown in Figs. \ref{fig:6}-\ref{fig:8}, and we found that the periodic signals $S_{1}(t)$, $S_{2}(t)$ and $S_{3}(t)$ can be clearly identified, respectively. The red dotted curves in Figs. \ref{fig:6}-\ref{fig:8} denote the original signals $S_{i}(t)$ ($i$ = 1, 2, 3) for comparison purpose.

\begin{figure}[hbt]
	\subfigure[]{
		\includegraphics[width=0.47\textwidth]{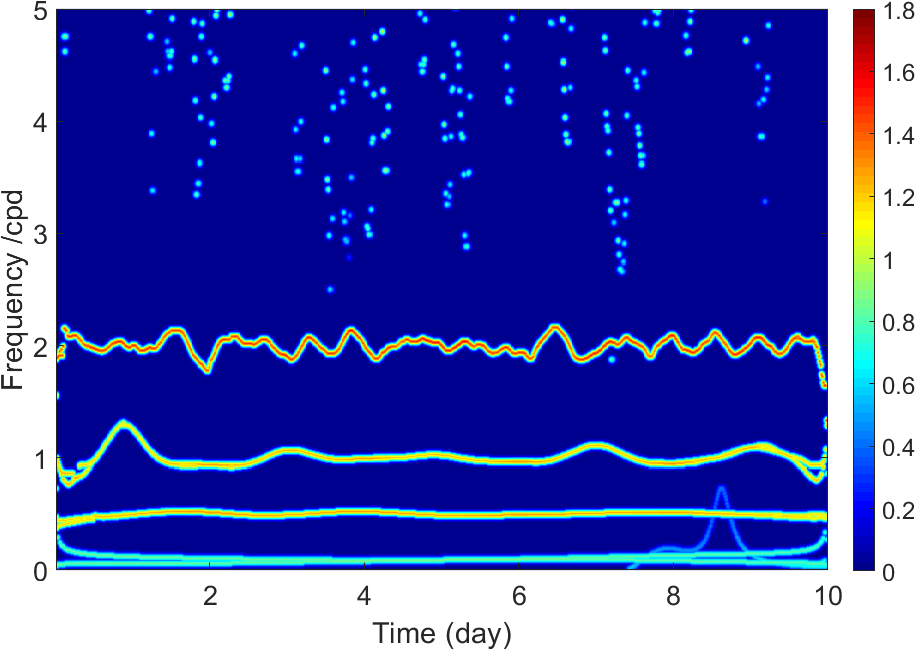} 
	}
	\quad
	\subfigure[]{
		\includegraphics[width=0.47\textwidth]{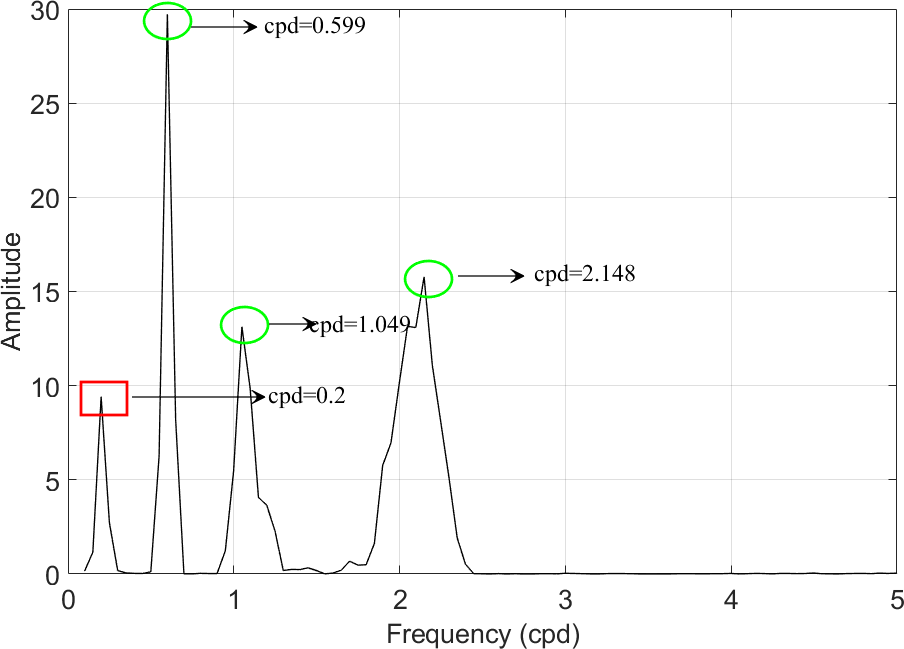} 
	}
	\caption{The frequency variations of each IMF (expect for trend r) decomposed from S(t) in Case 1 (with 2$\%$ of the magnitude of the noise). (a) The Hilbert spectrum of IMFs, where the color variations occurred in each skeleton curve represent the corresponding energy variations of each IMF, and (b) the corresponding marginal spectrum of IMFs.
	}
	\label{fig:9} 
\end{figure}

\begin{figure}[hbt]
	\subfigure[]{
		\includegraphics[width=0.47\textwidth]{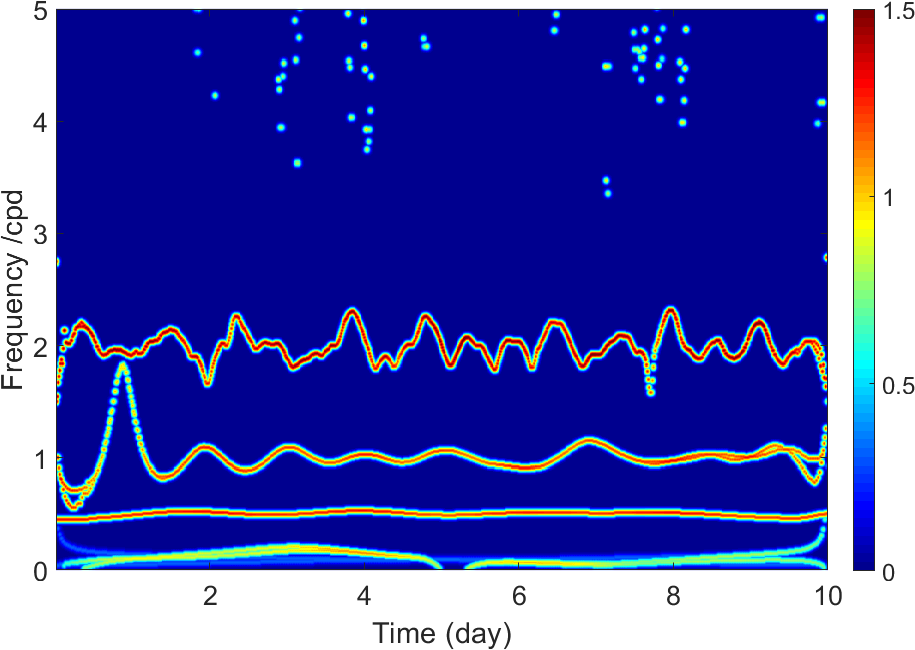} 
	}
	\quad
	\subfigure[]{
		\includegraphics[width=0.47\textwidth]{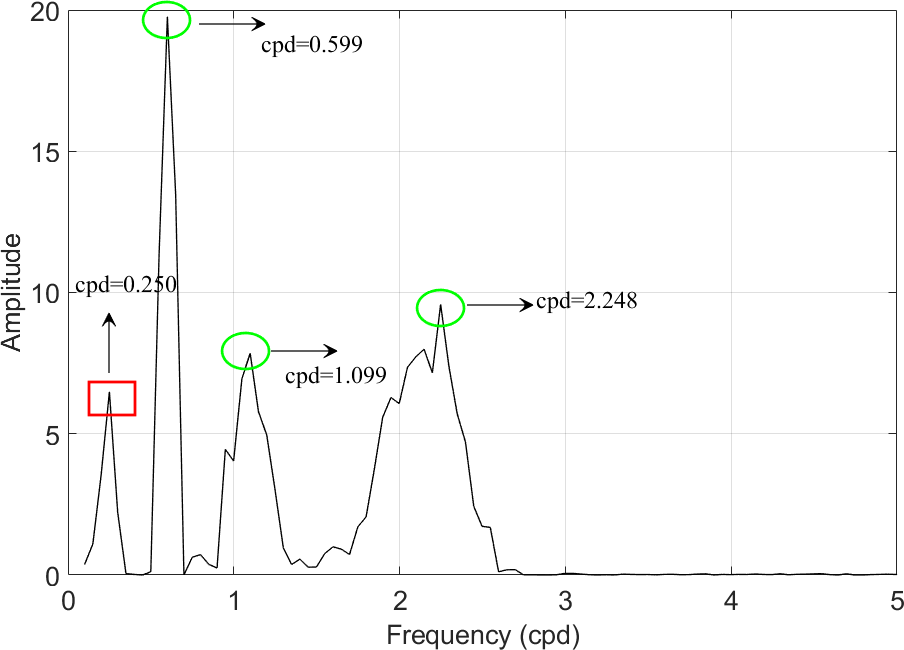} 
	}
	\caption{The frequency variations of each IMF (expect for trend r) decomposed from S(t) in Case 2 (with 5$\%$ of the magnitude of the noise). (a) The Hilbert spectrum of IMFs, where the color variations occurred in each skeleton curve represent the corresponding energy variations of each IMF, and (b) the corresponding marginal spectrum of IMFs.
	}
	\label{fig:10} 
\end{figure}

\begin{figure}[hbt]
	\subfigure[]{
		\includegraphics[width=0.47\textwidth]{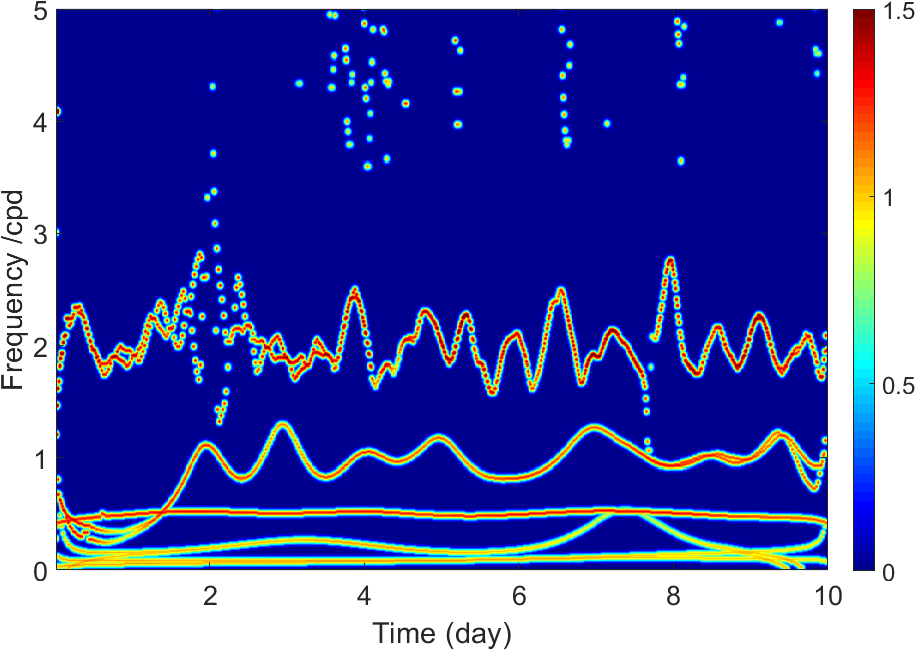} 
	}
	\quad
	\subfigure[]{
		\includegraphics[width=0.47\textwidth]{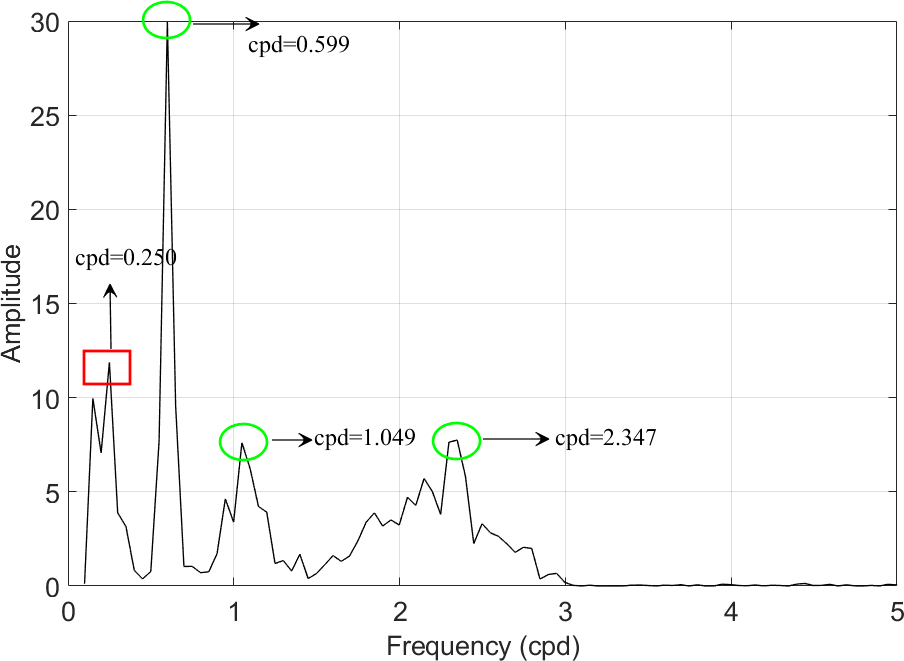} 
	}
	\caption{The frequency variations of each IMF (expect for trend r) decomposed from S(t) in Case 3 (with 10$\%$ of the magnitude of the noise). (a) The Hilbert spectrum of IMFs, where the color variations occurred in each skeleton curve represent the corresponding energy variations of each IMF, and (b) the corresponding marginal spectrum of IMFs.
	}
	\label{fig:11} 
\end{figure}

Further, the Hilbert transform (HT) is executed to display the variety of instantaneous frequencies for each IMF component. The corresponding Hilbert spectra and marginal spectra for IMFs (expect for r) in three cases are shown in  Figs. \ref{fig:9}-\ref{fig:11}, where each of the subfigures (a)s shows the frequency variations of each IMF, and each of the subfigures (b)s shows measures of the total amplitude (or energy) contribution from each frequency value, representing the cumulated amplitude over the entire data span in a probabilistic sense \citep{Huang1998Huang}. 

From Figs. \ref{fig:9}-\ref{fig:11}, we see that the mode-mixing problem exists in EEMD decomposition, and it becomes more obvious as noise increases from 2$\%$ to 10$\%$. However, the three periodic signals $S_{1}(t)$, $S_{2}(t)$, and $S_{3}(t)$ can be identified clearly. The set frequencies of $S_{1}(t)$, $S_{2}(t)$, and $S_{3}(t)$ are 2 circle per day (cpd), 1 cpd and 0.5 cpd, respectively, denoted as $f_{1-set}=2$ cpd, $f_{2-set}=2$ cpd, and $f_{3-set}=0.5$ cpd. After EEMD decomposition, the marginal spectra show that the corresponding values are $f_{1-case1}=2.148$ cpd, $f_{2-case1}=1.049$ cpd, $f_{3-case1}=0.599$ cpd (Case 1); $f_{1-case2}=2.248$ cpd, $f_{2-case2}=1.099$ cpd, $f_{3-case2}=0.599$ cpd (Case 2); $f_{1-case3}=2.347$ cpd, $f_{2-case3}=1.049$ cpd, $f_{3-case3}=0.599$ cpd (Case 3), respectively. And the detected signals corresponding to the original set signals $S_{1}(t)$, $S_{2}(t)$ and $S_{3}(t)$ are shown by the peaks denoted in green circles. In addition, there appear not-real signals with frequencies around 0.2 cpd after EEMD decomposition, which are denoted as $F(t)$ and shown by the peaks denoted in red rectangle. The  $F(t)$  might be meaningless in physical explaining, and it will be an interference if we focus on periodic signals. In this study, however, the target is to detect and identify the linear signal $S_{4}(t)$, which means that all periodic signals are useless and should be removed. After all these periodic signals are removed, we reconstruct a new signal series by summing the residual IMFs and the residual trend r. The reconstructed signal series is denoted as $S^{'}(t)$.

\begin{figure}[hbt]
	\centering
	\subfigure[]{
	\includegraphics[width=0.7\textwidth]{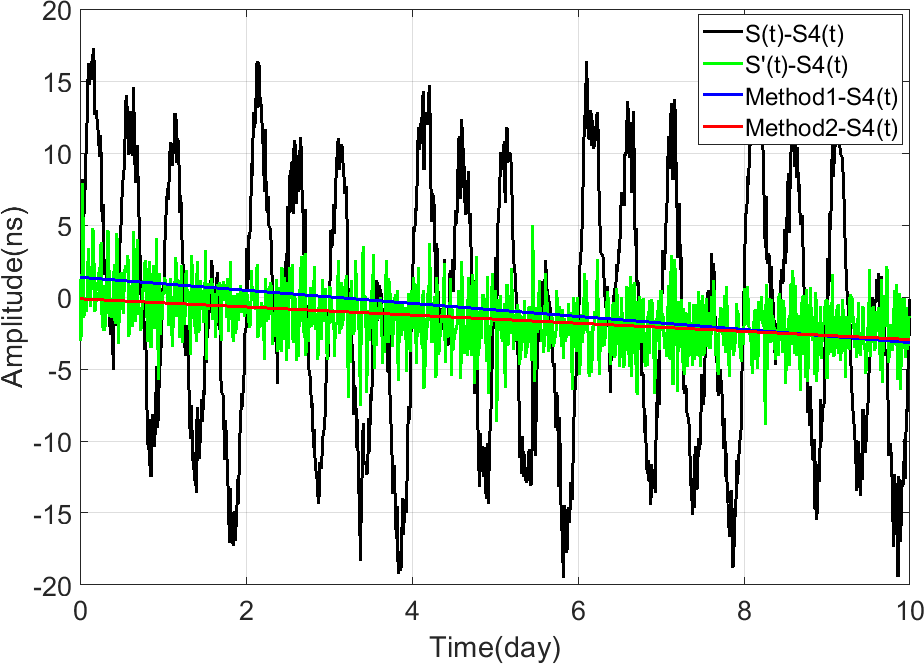} 
}
	\subfigure[]{
	\includegraphics[width=0.7\textwidth]{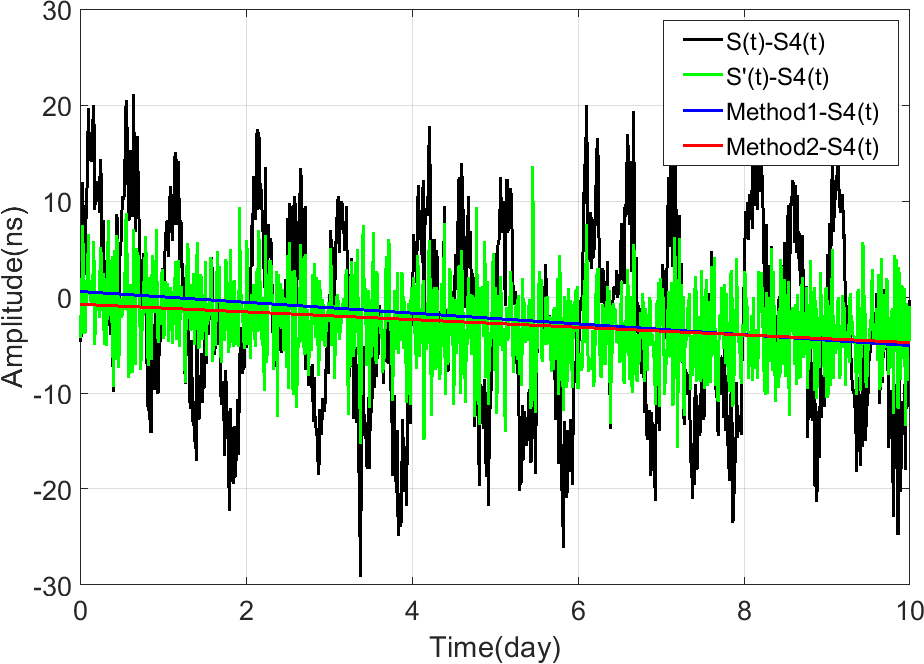} 
}
	\subfigure[]{
	\includegraphics[width=0.7\textwidth]{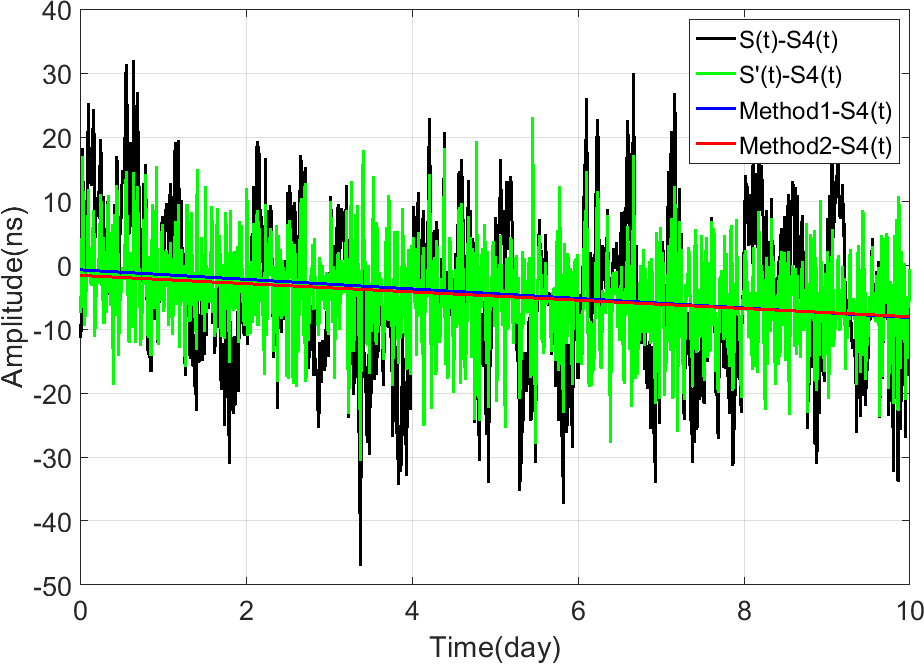} 
}
	\caption{The comparison of the linear fitting for the signal series $S(t)$ (Method1) and the linear fitting for the reconstructed signal series $S^{'}(t)$ (Method2). $S_{4}(t)$ is a given real signal series. (a), (b) and (c) denote the cases of noise magnitudes of 2$\%$, 5$\%$ and 10$\%$ of $S_{base}(t)$, respectively.}
	
	\label{fig:12} 
\end{figure}

We take the signal series $S_{4}(t)$ as real signal series, and try to recovery it by two different methods. The first method is to perform a least squares linear fitting on the signal series $S(t)$  directly; and the second method is to perform EEMD decomposition on $S(t)$ and then reconstruct the new signal series $S^{'}(t)$ by removing periodic series IMFs, and the least squares linear fitting is performed on $S^{'}(t)$. The results are shown in Fig. \ref{fig:12}.

Here we use the standard deviation (STD) of the difference series between the real signal series $S_{4}(t)$ and that determined by Method 1 or 2 to evaluate the reliability of the two methods. Using Method 1, namely the direct least squared linear fitting of the series $S(t)$, the STDs of the results are 1.31 ns (Case 1), 1.63 ns (Case 2), and 2.16 ns (Case 3), respectively. Using Method 2, namely the least squared linear fitting of the reconstructed series $S^{'}(t)$, which is obtained after removing the periodic series via EEMD technique, the STDs of the results are 0.82 ns (Case 1), 1.16 ns (Case 2), and 1.87 ns (Case 3), respectively. The comparative results clearly suggest that the EEMD technique is effective for extracting the linear signals of interest by a priori removing the contaminated periodic signals from the original observations.

\section{Experiments and data processing}
\label{sec: 5}
\subsection{Experiments setup}
\label{sec:  5.1}
\begin{figure}[hbt]
	\centering
	\includegraphics[width=0.8\textwidth]{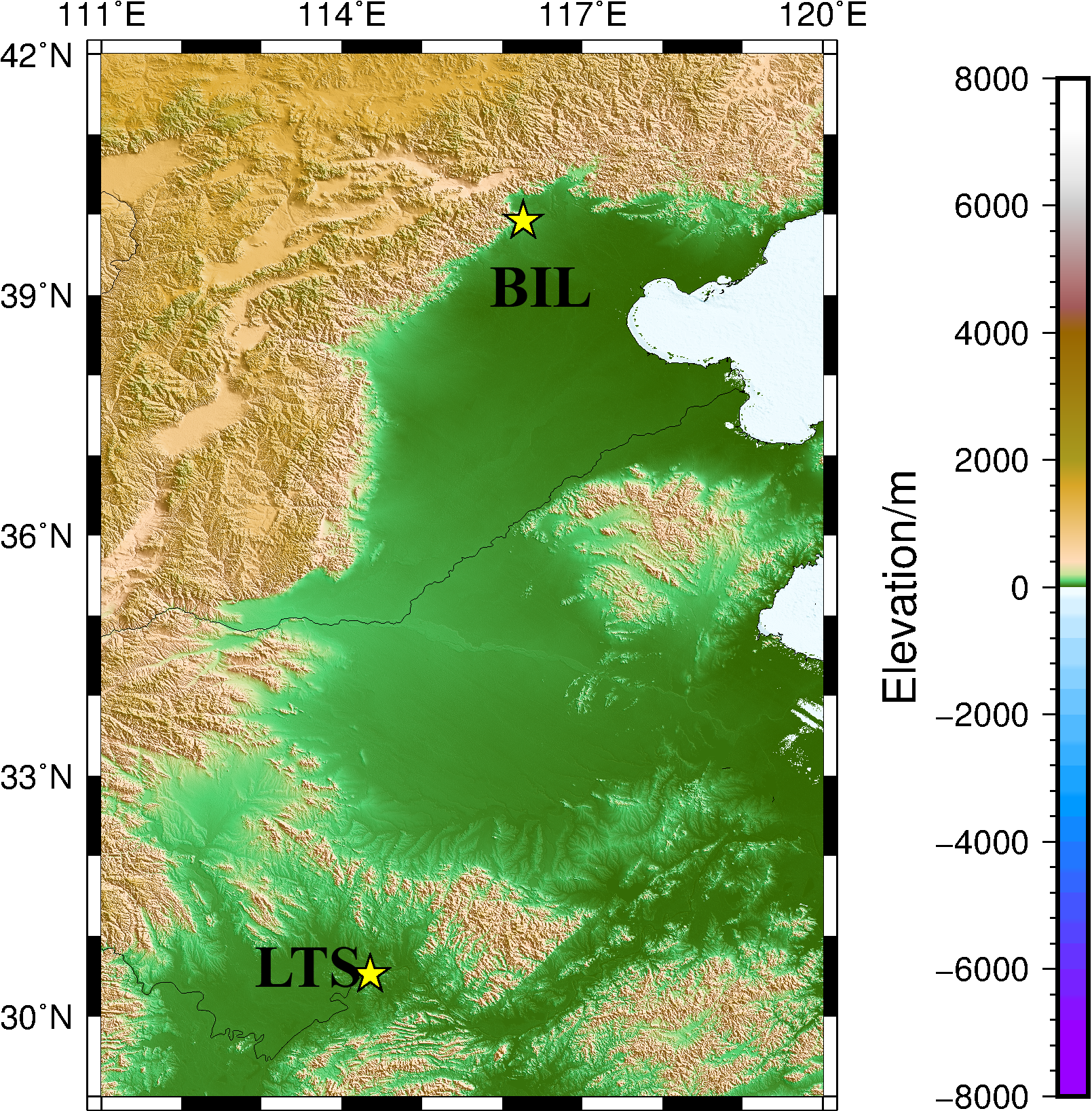}
	\caption{The distribution of the two ground stations of BIL and LTS in the experiments. The two stations are around 1000 km away; the OH difference of the antennas and clocks ($H_{clock_{_A}}-H_{clock_{_B}}$) between BIL and LTS are 37.3 m and 16 m, respectively.}
	\label{fig:13}
\end{figure}

In our experiments, we used two hydrogen time and frequency standards, with one is a fixed reference clock $C_{A}$ (iMaser3000) and the other is a portable clock $C_{B}$ (BM2101-02). The experiments are conducted at Beijing 203 Institute Laboratory (BIL) and Luojiashan Time-Frequency Station (LTS), and the locations of the two stations are shown in Fig. \ref{fig:13} and Table \ref{tab:2}. And we used the Modified Allan deviation (MDEV)\citep{Allan1991A} to evaluate the frequency stability of the two hydrogen atomic clocks, which is estimated from a set of frequency measurements for time averaging. The MDEV is expressed as follows \citep{Allan1991A,Bregni1997Estimation,Lesage2007Characterization}:
\begin{equation}\label{eq:34}
\delta_{_M}(\tau)=\sqrt{\frac{\sum_{j=1}^{N-3n+1}[\sum_{i=j}^{n+j-1}(x_{i+2n}-2x_{i+n}+x_{i})]^{2}}{2\tau^{2}n^{2}(N-3n+1)}}
\end{equation}
where $N$ denotes the total number of the time series, $\tau_{0}$ denotes the basic time interval, and $\tau=n\tau_{0}$.

The frequency stabilities of the two clocks on the different average time intervals are given in Table \ref{tab:3}. The MDEV with shorter interval represents random noise of the clock, while with longer interval represents secular frequency drift \citep{Kopeikin2016Chronometric}.
\begin{table*}
	\caption{The detailed information of the two ground stations (BIL and LTS) in the transportation experiments. The coordinates ($\varphi, \lambda, h$) denote the locations of GNSS antennas (under the frame of WGS 84); $H_{ant}$ and $H_{clock}$ denote the OHs of GNSS antenna and corresponding location of hydrogen atomic clock $C_{i}$, respectively. It should be noted that the OH of GNSS antenna, $H_{ant}$, is determined by EGM2008 gravity field model. And the OH of clock $C_{i}$, $H_{clock}$, is determined by measuring the OH difference via a ruler.}
	\label{tab:2}
	\begin{center}
		\setlength{\tabcolsep}{8.5mm}{
			\begin{tabular}{lllllll}
				\hline\noalign{\smallskip}
				Stations & $\varphi (\circ)$ & $\lambda (\circ)$ & $h(m)$ & $H_{ant}(m)$ & $H_{clock}(m)$ \\
				\noalign{\smallskip}\hline\noalign{\smallskip}
				BIL &116.26 & 39.91 & $70.51$ &$79.09$ & $51.55$ \\
				LTS &114.36 & 30.53 & $28.00$ &$41.76$ & $35.51$ \\
				\noalign{\smallskip}\hline
		\end{tabular}}
	\end{center}
\end{table*}

\begin{table*}
	\caption{The Modified Allan Deviation (MDEV) of the two hydrogen time and frequency standards (the fixed one: iMaser3000($C_{A}$) and the portable one: BM2101-02($C_{B}$)).}
	\label{tab:3}
	\begin{center}
		\setlength{\tabcolsep}{4.5mm}{
			\begin{tabular}{lllllll}
				\hline\noalign{\smallskip}
				Time interval & 1 s & 10 s & 100 s & 1000 s & 10000 s  \\
				\noalign{\smallskip}\hline\noalign{\smallskip}
				$C_{A}$ & $1.5 \times 10^{^{-13}}$ & $2.0 \times 10^{^{-14}}$ & $5.0 \times 10^{^{-15}}$ & $2.0 \times 10^{^{-15}}$ & $2.0 \times 10^{^{-15}}$ \\
				$C_{B}$ & $4.57 \times 10^{^{-13}}$ & $8.85 \times 10^{^{-14}}$ & $1.95 \times 10^{^{-14}}$ & $5.96 \times 10^{^{-15}}$ & $2.18 \times 10^{^{-15}}$ \\
				\noalign{\smallskip}\hline
		\end{tabular}}
	\end{center}
\end{table*}

\begin{figure*}[hbt]
\centering
\includegraphics[width=1\textwidth]{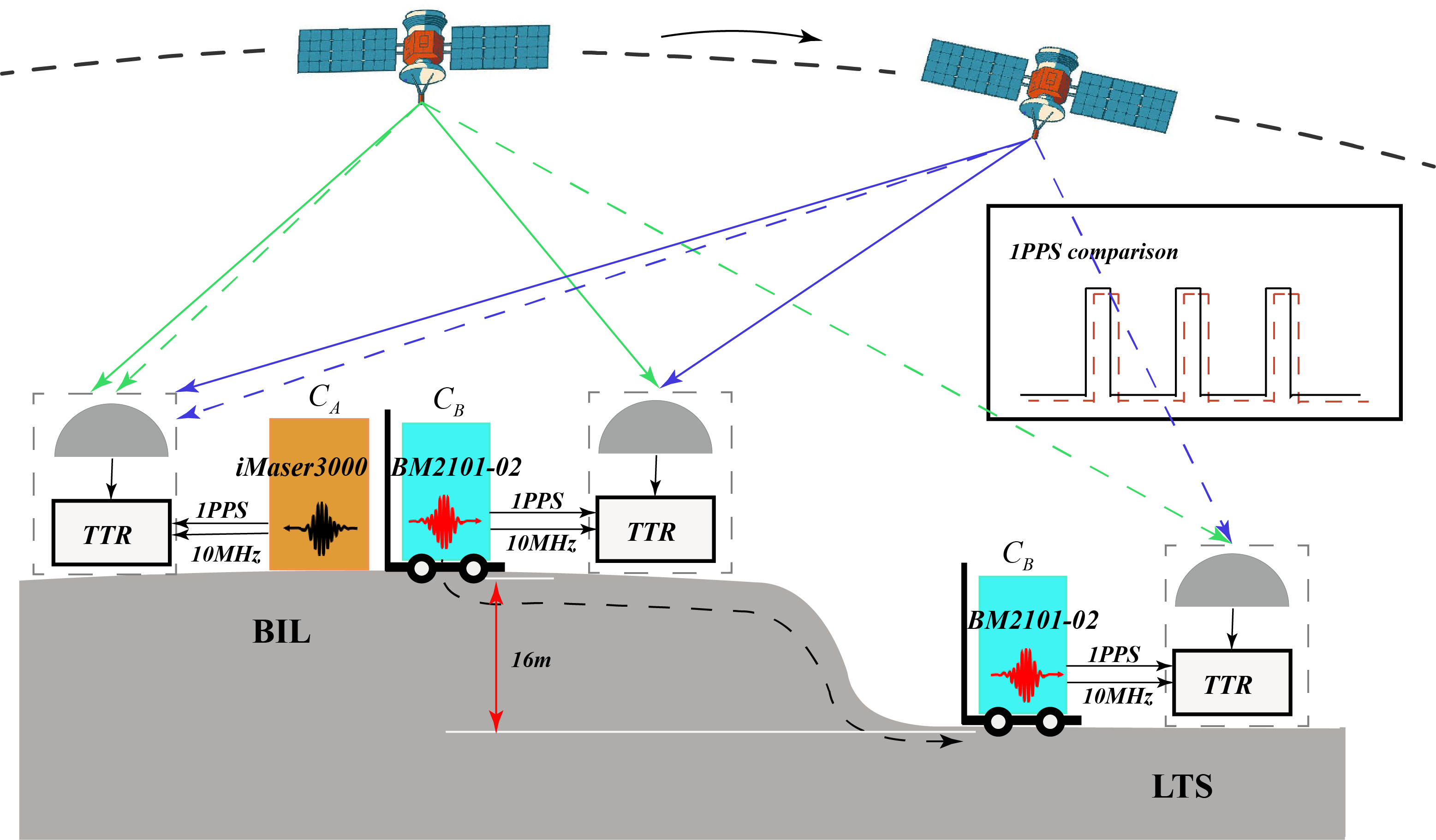}
\caption{The schematic diagram of the transportation experiments. In Period 1, the zero-baseline measurement is implemented with both $C_{A}$ and $C_{B}$ located at the BIL, and this period lasts from January 09, 2018 to January 29, 2018. After that the clock $C_{B}$ is transported to LTS from BIL, and the geopotential difference measurement is implemented in Period 2 with $C_{A}$ and $C_{B}$ located at BIL and LTS, respectively.The Period 2 lasts from Febrary 01, 2018 to April 07, 2018.}
\label{fig:14}
\end{figure*}

The experiments are divided into two Periods (shown in Fig. \ref{fig:14}). In Period 1 that spans from January 09, 2018 to January 29, 2018, the zero-baseline measurement is implenented at BIL with clocks $C_{A}$ and $C_{B}$ are placed in the same shielding room with the same geopotentials. According to the International GNSS Tracking Schedule \citep{Allan1994Technical,Defraigne2015CGGTTS}, the clock comparisons between $C_{A}$ and $ C_{B}$, $\Delta C_{AB}$, are conducted via CVSTT technique, and a series of clock offsets $\Delta C_{AB}(t)$ can be obtained, where $\Delta C_{AB}=C_{B}-C_{A}$, and $\Delta C_{AB}(t)=C_{B}(t)-C_{A}(t)$.

When zero-baseline measurement is finished, the portable clock $C_{B}$ is transported to LTS from BIL while keeps clock $C_{A}$ unmoved at the BIL, and during transportation the portable clock $C_{B}$ continues operating. After installation and adjustment of $C_{B}$ at the LTS, the clock comparisons between  $C_{A}$ and $C_{B}$ are conducted in geopotential difference measurement, similarly. And this period is denoted as Period 2, which spans from Febrary 01, 2018 to April 07, 2018. It should be noted that there is no difference except for the location diversity of clock $C_{B}$, when comparing the setup difference of Period1 and Period 2. In addition, there are two days of missing data (from January 30, 2018 to January 31, 2018) due to the fact that during transportation (around 8 h) there are no observations and the observations from the day after the re-observation are not stable and consequently not used.

Due to the fact that temperature is a major factor which may disturb the performance of atomic clocks as well as the cable delays \citep{Weiss1999Effects, Lisowiec2004Progress},we controll the environment with a relatively constant temperature in whole experiments. The temperature is held nearly constant ( $ 24\pm0.5 ^{\circ}C$) for the fixed clock $C_{A}$ in Period 1 and Period 2 at BIL. For the portable clock $C_{B}$ , it is in the same laboratory with clock $C_{A}$ during Period 1, and the temperature is locked at $24 \pm1.5 ^{\circ}C$ in Period 2 at LTS.

\subsection{ Data Processing}
\label{sec: 5.2}
\begin{figure}[hbt]
	\centering
	\includegraphics[width=1\textwidth]{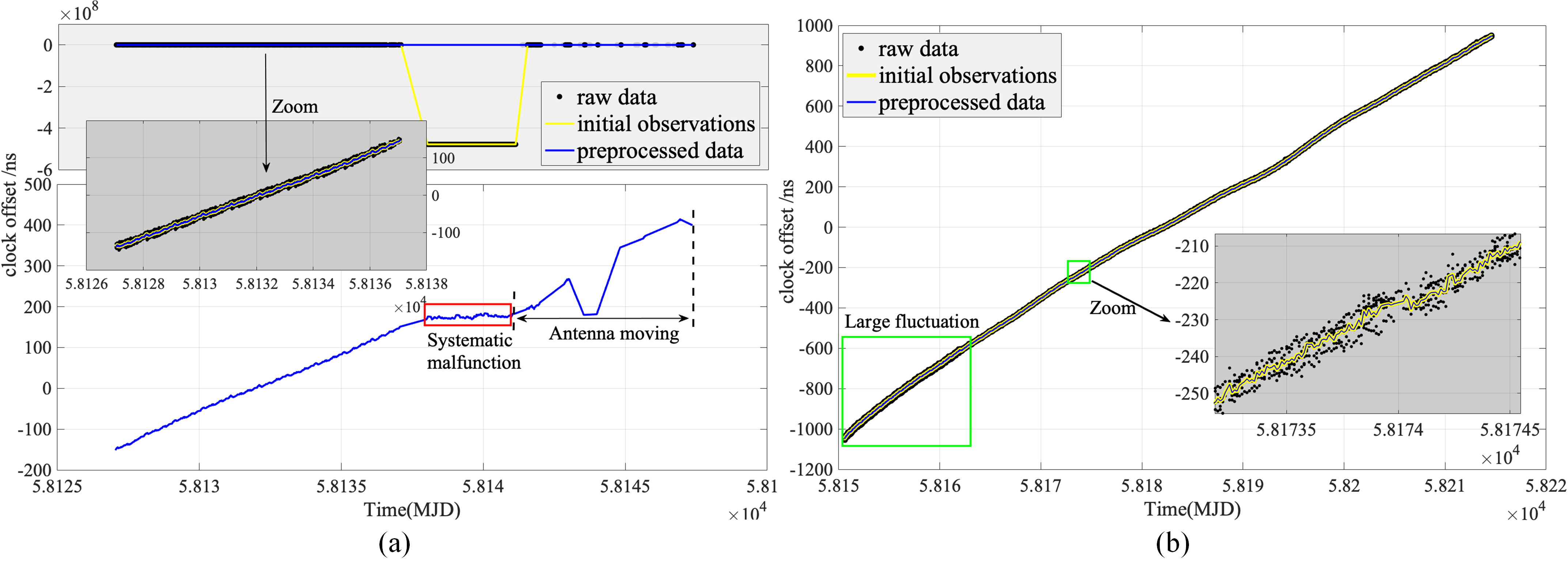}
	\caption{The data preprocessing for clock offsets series $\Delta C_{AB}(t)$ via the CVSTT technique. (a) The data preprocessing of the zero-baseline measurement, (b) the data preprocessing of the geopotential difference measurement. The raw data (black dotted curve) denotes the results of $\Delta C_{AB}(t)$ from all common-view satellites; the initial observations (yellow curve) denotes the results of $\Delta C_{AB}(t)$ which is the weighted average based on satellite elevations; the preprocessed data (blue curve) denotes the results of $\Delta C_{AB}(t)$ after data preprocessing on initial observations (after gross error removing, clock jump correction, and data insertion).}
	\label{fig:15}
\end{figure}

Based on the observations via CVSTT technique, the clock offsets series $\Delta C_{AB}(t)$ in each Period are determined. Here, the clock offsets series $\Delta C_{AB}(t)$ of the observations for all common-view satellites during Period 1 and Period 2 are denoted by the black dotted curve (simply denoted as raw data) in Fig. \ref{fig:15}. And take the satellite elevation angles into consideration, the initial CVSTT observations (initial observations) are obtained, which are denoted by the yellow curve in Fig. \ref{fig:15}. After that, we performe data preprocessing on the initial observations by removing gross error, correcting clock jumps and inserting missing data, and the preprocessed CVSTT data (preprocessed data) are finally obtained which denoted by the blue curve in Fig. \ref{fig:15}. After data preprocessing, the clock offsets series $\Delta C_{AB}(t)$ become more stable. In the following data processing, all operations are based on the preprocessed data sets.

\begin{figure}[hbt]
	\centering
    \includegraphics[width=1\textwidth]{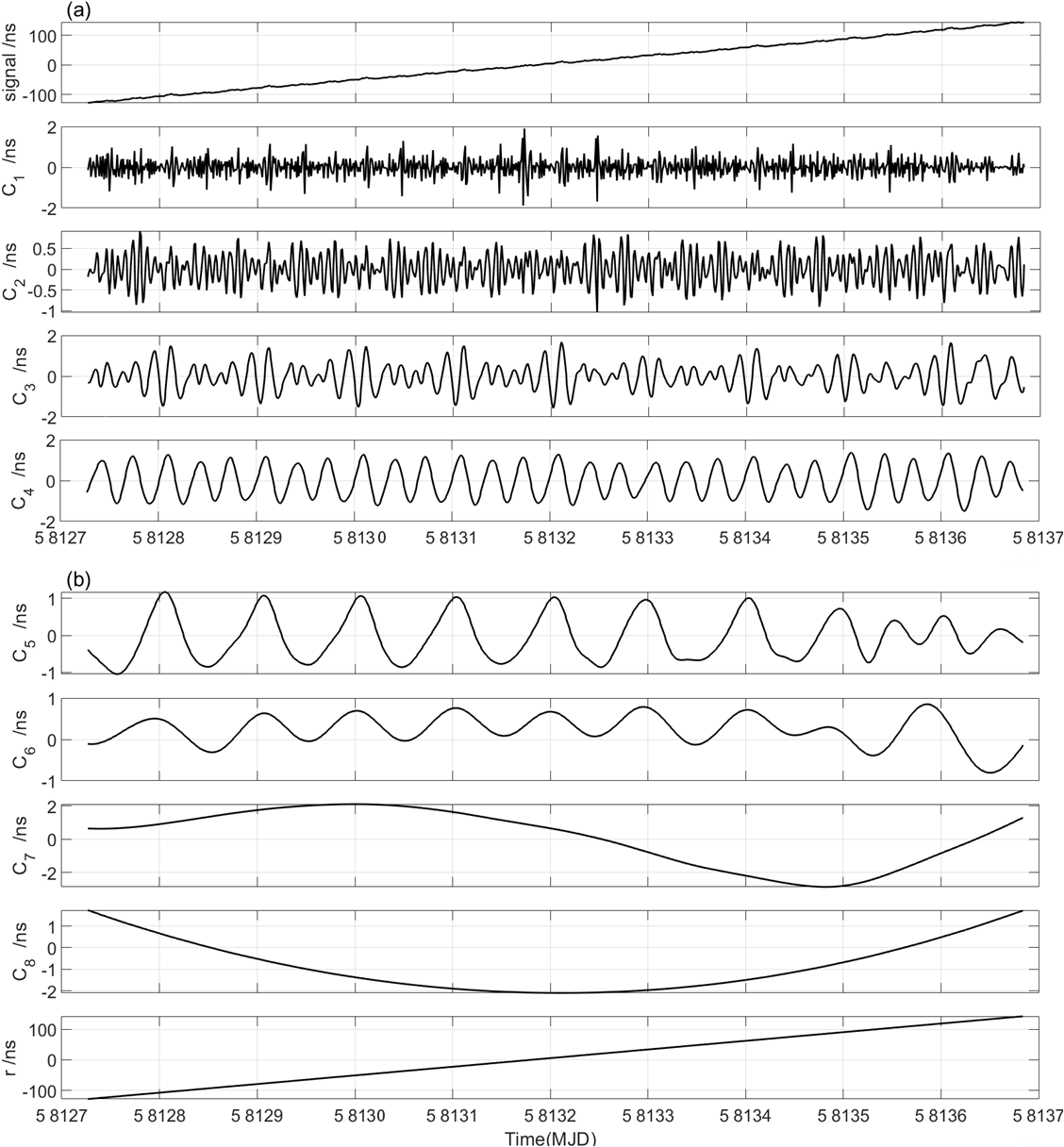}
	\caption{The resulting EEMD components of clock offsets series $\Delta C_{AB}(t)$ in zero-baseline measurement (Period 1) with both $C_{A}$ and $C_{B}$ located at BIL, which lasts from MJD 58127 to MJD 58136. (a) The preprocessed data and the components IMF1 $\thicksim$ IMF4, (b) the components IMF5 $\thicksim$ IMF8 and the residual trend $r$.}
	\label{fig:16}
\end{figure}

At the same time, it should be noted that, however, the observations from MJD 58137 to MJD 57141 are not used due to a systematic malfunction, and the observations from MJD 58142 to MJD 57147 are also not used due to random movement of the GNSS antenna (the two clocks are not transported, but one antenna is replaced). Therefore, the valid observations in Period 1 spans from MJD 58127 to MJD 57136 (seen in Fig. \ref{fig:15}(a)). In Period 2, since the stability of clock $C_{B}$ could be significantly influenced by various environment factors during its transportation to LTS from BIL, the initial observations at LTS contains large fluctuations and noises (from MJD 58150 to 58161, seen in Fig. \ref{fig:15}(b) and Fig. \ref{fig:20}(b)), which will contaminate final determination of the geopotentials. Hence, as for preprocessed data in Period 2, we adopt two different preprocessed data sets by whether contains the observations of the first 12 days, and the details are shown later.

\begin{figure}[hbt]
\centering
\includegraphics[width=1\textwidth]{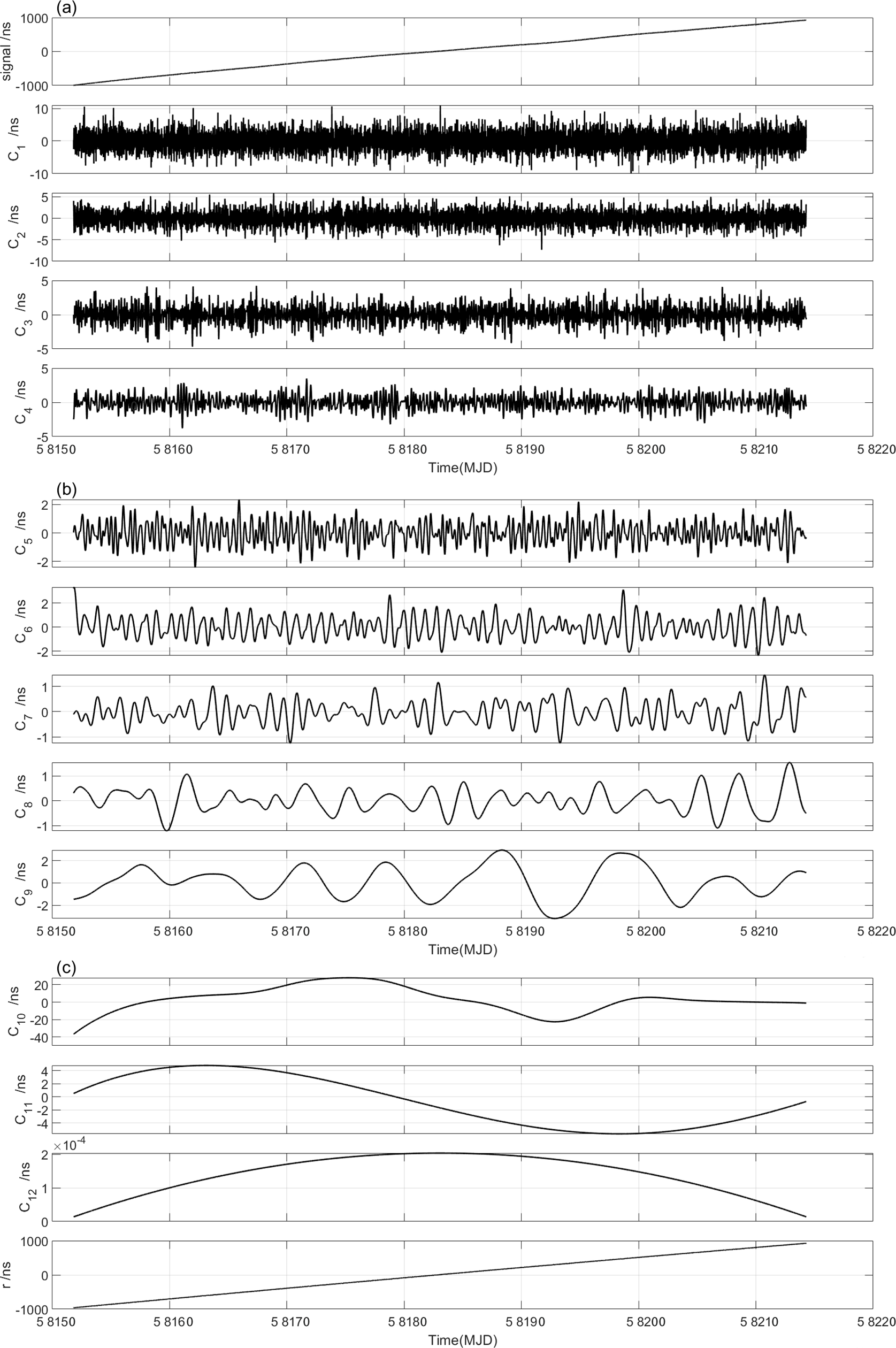}
\caption{The resulting EEMD components of clock offsets series $\Delta C_{AB}(t)$ in geopotential difference measurement (Period 2) with $C_{A}$ located at BIL and $C_{B}$ located at LTS, which lasts from MJD 58150 to MJD 58215. (a) The preprocessed data and the components IMF1 $\thicksim$ IMF4, (b) the components IMF5 $\thicksim$ IMF9, (c) the components IMF10 $ \thicksim$ IMF13 and the residual trend $r$.}
\label{fig:17}
\end{figure}

Then, the  EEMD technique is applied to the preprocessed data sets for removing the uninteresting periodic components and extracting the geopotential-related signals. The preprocessed data sets in Period 1 and Period 2 are decomposed into a series of IMFs (with frequencies from higher to lower) and a long trend component $r$ after EEMD decomposition, as shown in Figs. \ref{fig:16} and \ref{fig:17}. Generally, the components of the EEMD convey a physical meaning due to the fact that the characteristic scales are physical \citep{Huang1998Huang}. However, the first several high frequency components might be fictitious due to the fact that the sampling interval (960s) is too large to capture the high-frequency variations. As a result, the data are jagged at these frequencies (e.g. $c_{1}$ in Fig. \ref{fig:16}; $c_{1}$, $c_{2}$, $c_{3}$ in Fig. \ref{fig:17}).

\begin{figure}[hbt]
	\centering
	\includegraphics[width=1\textwidth]{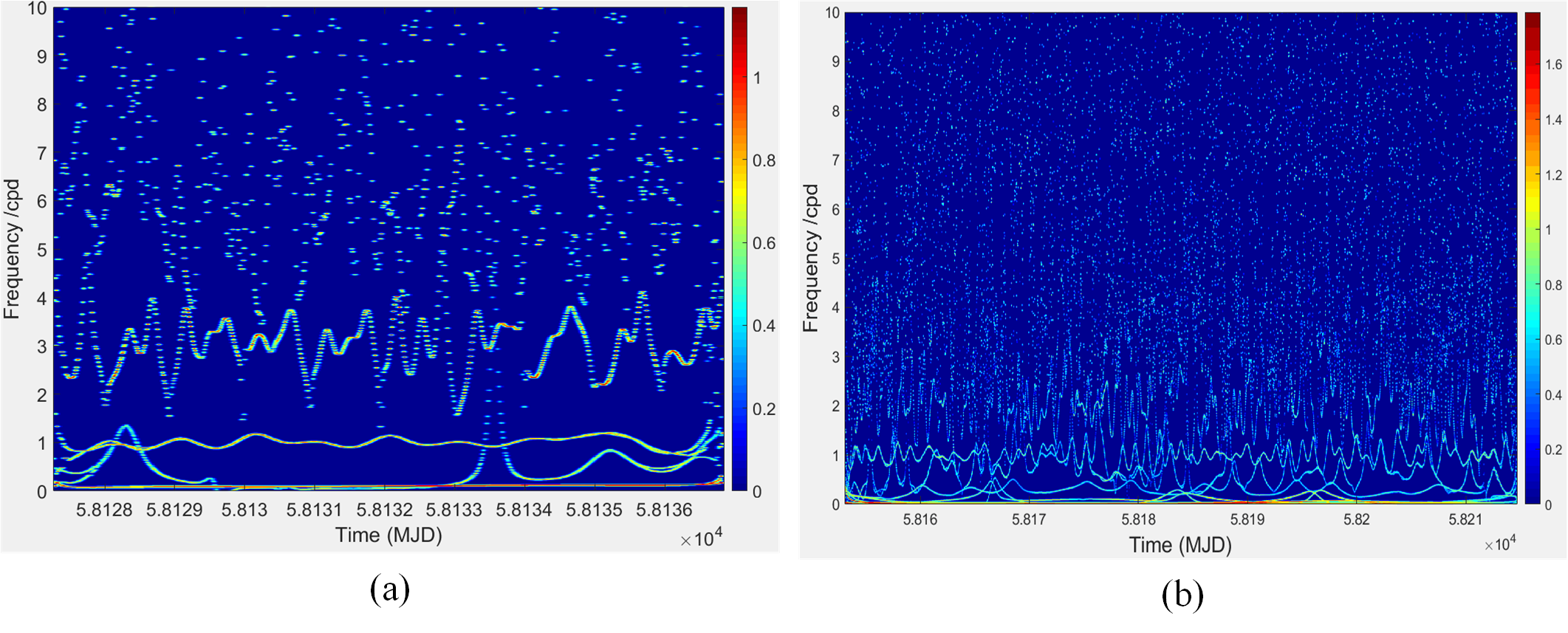}
	\caption{The corresponding Hilbert spectra of EEMD decomposition in zero-baseline measurement (a) and that of in geopotential difference measurement (b).}
	\label{fig:18}
\end{figure}

\begin{figure}[hbt]
	\centering
	\includegraphics[width=1\textwidth]{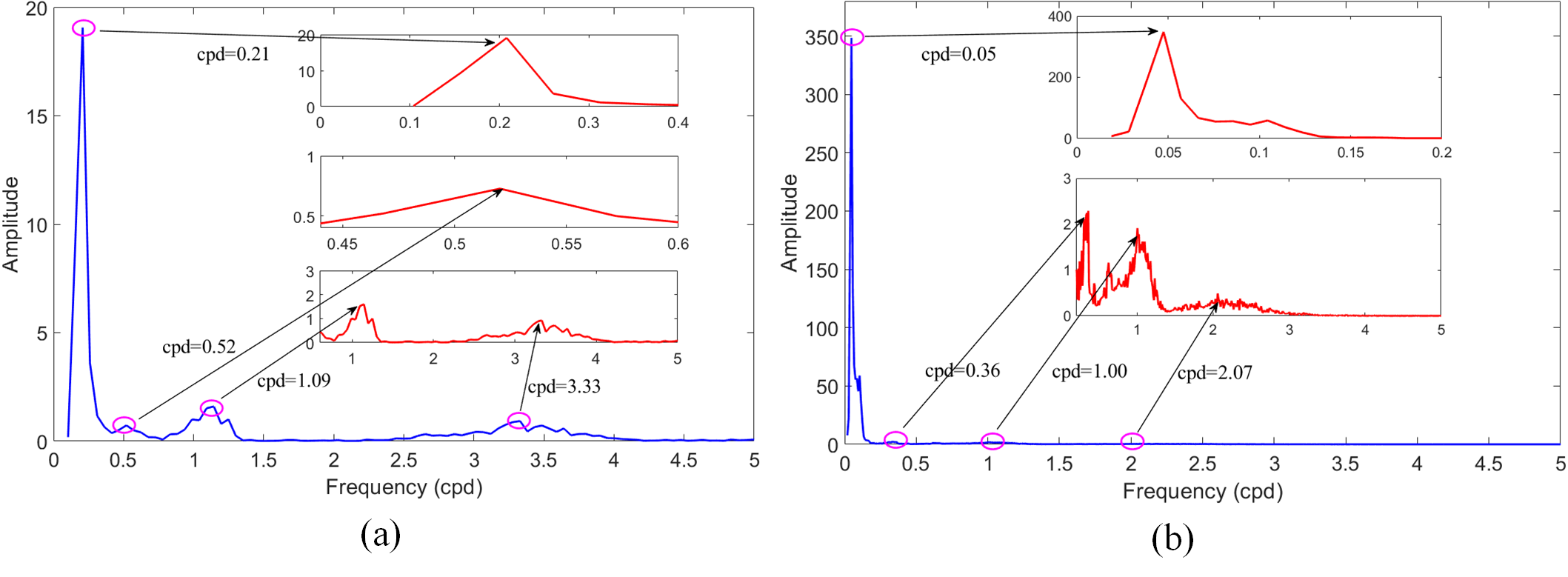}
	\caption{The corresponding marginal spectra of EEMD decomposition in zero-baseline measurement (a) and that of in geopotential difference measurement (b).}
		\label{fig:19}
	\end{figure}

After using EEMD technique for preprocessed data sets in Period 1 and Period 2, we examine completeness and orthogonality of the EEMD decomposition. The IO values corresponding to Period 1 and Period 2 are determined with values of $-0.0085$ and $-0.0094$, respectively, demonstrating that the signals are effectively decomposed. It should be noted that the IO value during Period 2 is worse than that in Period 1. This might be due to the environmental difference of $C_{B}$ and $C_{A}$ during Period 2, or due to the difference of the signal propagation paths during Period 2.

The corresponding Hilbert spectra in Period 1 and Period 2 are given in Fig. \ref{fig:18}. In order to measure the energy contributions from each frequency value, the marginal spectra of all IMFs are given in Fig. \ref{fig:19}, which  represent the cumulated amplitude over the entire data span in a probabilistic sense. From Figs. \ref{fig:18} and \ref{fig:19}, the periodic signals with frequencies of around 0.2 cpd, 0.5 cpd, 1 cpd and 3.3 cpd of zero-baseline measurement are detected; and the periodic signals with frequencies of around 0.05 cpd, 0.36 cpd, 1 cpd and 2 cpd of geopotential difference measurement are detected. Finally, these periodic signals included in the original CVSTT observation series are removed, and we reconstructed the residual clock offsets series $\Delta C_{AB-re}(t)$ by summing the residual components. The reconstructed clock offsets series $\Delta C_{AB-re}(t)$ are regarded as the geopotential-related signals, based on which we can determine the geopotential difference between $C_{B}$ (LTS) and $C_{A}$ (BIL). The reconstructed clock offsets series $\Delta C_{AB-re}(t)$ are denoted as EEMD results,  as described in section \ref{sec: 4.2}.

\section{Results}
\label{sec:6}

\begin{figure}[hbt]
	\centering
	\includegraphics[width=1\textwidth]{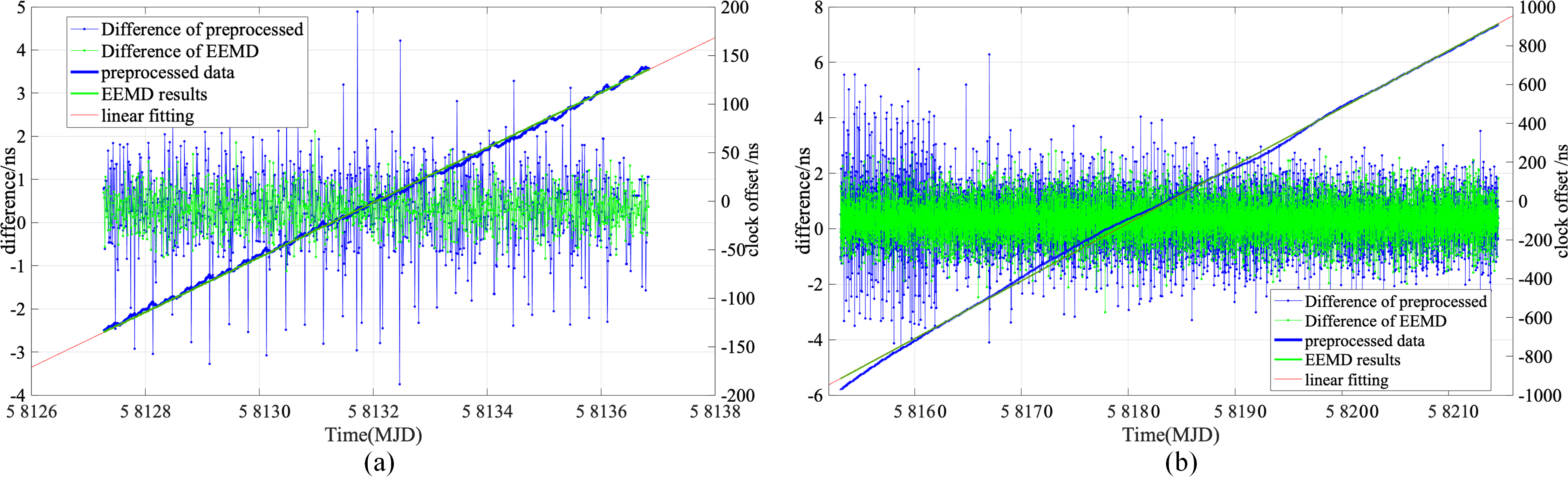}
	\caption{The experimental results of clock offsets series via CVSTT technique. (a) The clock offsets series of zero-baseline measurement (Period 1), (b) the clock offsets series of geopotential difference measurement (Period 2). It should be noted that the blue curve is the preprocessed data sets $\Delta C_{AB}(t)$; the green curve is the EEMD results $\Delta C_{AB-re}(t)$, which is determined after EEMD technique; the red curve is a linear fitting for the $\Delta C_{AB-re}(t)$.
	}
	\label{fig:20}
\end{figure}

After data processing, the corresponding EEMD results of clock offsets series $\Delta C_{AB-re}(t)$ in Period 1 and Period 2 are determined, as shown in Fig. \ref{fig:20}, which provides the clock offsets series before and after EEMD technique via the CVSTT technique. We can find that the clock offsets series become more stable and smooth after removing periodic signals included in the preprocessed data sets. Concerning zero-baseline measurement, after EEMD decomposition and removing periodic IMFs, the EEMD results of the reconstructed clock offsets $\Delta C_{AB-re}(t)$ are determined directly, and the corresponding time elapse difference, $\alpha_{zero}$, is then determined based on formula \eqref{eq:6}. We take this result as a system constant shift.

\begin{figure}[hbt]
	\centering
	\includegraphics[width=1\textwidth]{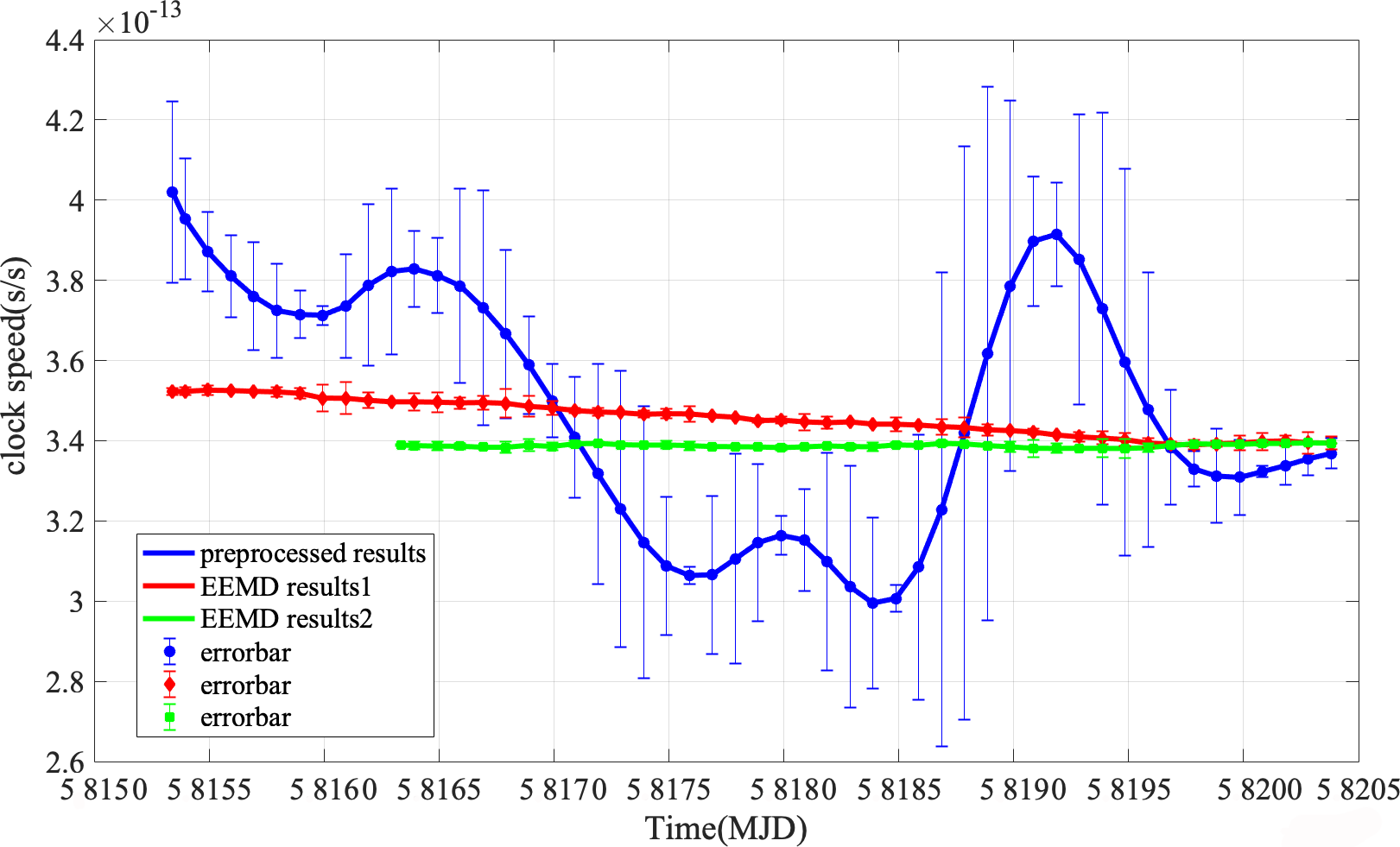}
	\caption{the variation of the time elapse difference $\alpha(t)$ ($\alpha(t)=\alpha_{geo}(t)-\alpha_{zero}$). The $\alpha(t)$ is caused by geopotential difference between the two staions ($LTS$ and $BIL$). The preprocessed results (blue curve) are determined based on the preprocessed data set; the EEMD results1 (red curve) are determined based on EEMD results which contains whole observations in Period 2; the EEMD results2 (green curve) are determined based on EEMD results which removes the first 12-day observations before EEMD technique.}
	\label{fig:21}
\end{figure}

For the geopotential difference measurement, however, the first 12-day observations contain large fluctuations and noises (seen in Fig. \ref{fig:20}b) when compared to  the later observations. Therefore, the preprocessed data set in Period 2 are analyzed twice. For the first analysis we used the entire data span to implement EEMD decomposition (the red curve in Fig. \ref{fig:21}), and for the second analysis we removed the first 12-day observations before EEMD decomposition for avoding data pollution (the green curve in Fig. \ref{fig:21}). In addition, for the geopotential difference measurement, we segment the corresponding $\Delta C_{AB-re}(t)$ into 10-day units, not only for matching the zero-baseline duration, but also for improving data utilization. Furthermore, by this strategy, the clock drift could be limited to a relative short time duration (10 days). And we use MDEV of each unit as the weighting factor for determining the final experimental results, which means that for a certain time duration, the higher the stability the greater the weight. By the above strategy, the interference of clock drift could be limited to a relative low level. And based on each unit, a series of time elapse differences $\alpha_{geo}(t)$ are determined.

\begin{figure}[hbt]
\centering
\includegraphics[width=1\textwidth]{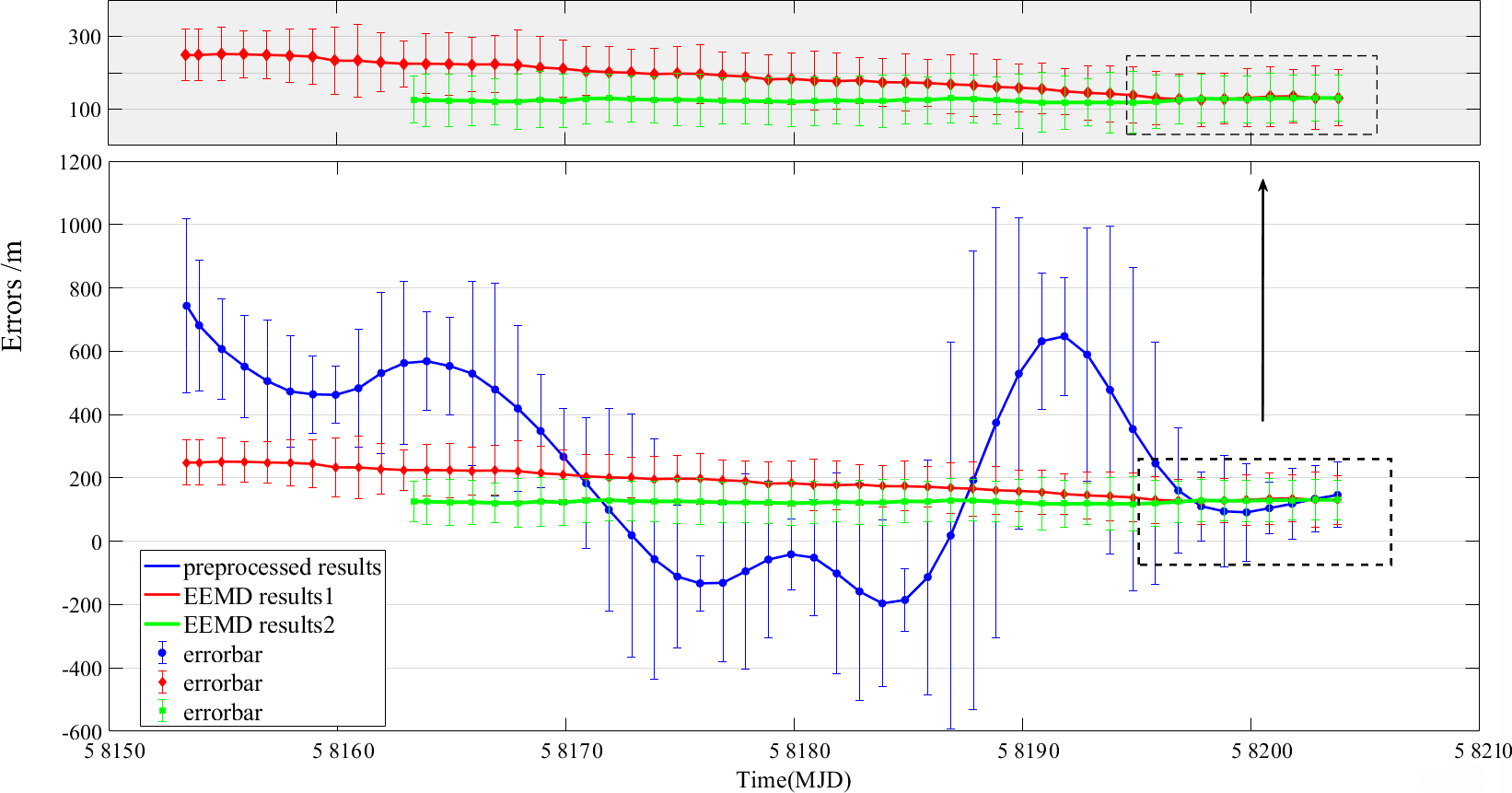}
\caption{The discrepancy between the clock-comparison-determined results and the corresponding EGM2008 results $D(t)$ ($D(t)=H_{_{LTS}}^{_{(T)}}(t)-H_{_{LTS}}$). The preprocessed results (blue curve) are determined based on the preprocessed data set; the EEMD results1 (red curve) are determined based on EEMD results which contains whole observations in Period 2; the EEMD results2 (green curve) are determined based on EEMD results which removes the first 12-day observations before EEMD technique.}
\label{fig:22}
\end{figure}

 According to equations \eqref{eq:2} $\thicksim$ \eqref{eq:6}, by taking the results of $\alpha_{zero}$ in Period 1 as system constant shift, the variation of the time elapse difference caused by the geopotentials, $\alpha(t)$ is finally determined with $\alpha(t)=\alpha_{geo}(t)-\alpha_{zero}$. The $\alpha(t)$ is caused by the geopotential difference between the two stations $LTS$ and $BIL$. Therefore, the clock-comparison-determined results of  geopotential difference $\Delta W_{_{AB}}^{_{(T)}}$ as well as the OH of $H_{_{LTS}}^{_{(T)}}$ is obtained. We note that to determine the OH of  the station LTS, the  OH of  the station BIL is a priori given. The results of $\alpha(t)$ and $D(t)$ are shown in Figs. \ref{fig:21} and \ref{fig:22}, respectively. It is noted that the $D(t)=H_{_{LTS}}^{_{(T)}}(t)-H_{_{LTS}}$ is the discrepancy between the  clock-comparison-determined results $H_{_{LTS}}^{_{(T)}}(t)$ and the corresponding EGM2008 results $H_{_{LTS}}$. From Figs. \ref{fig:21} and \ref{fig:22}, we can find that after removing these periodic signals the results become more stable besides smaller errorbars, which explain the validity of the EEMD technique in extracting the linear signals of interest. In addition, comparing the EEMD results in case 1 (EEMD Results 1) and case 2 (EEMD Results 2),  the interference caused by transportation from BIL to LTS is reduced by removing the first 12-days observations in Period 2.

The results of the clock-comparison-determined geopotential difference {\footnotesize{$\Delta W_{_{AB}}^{_{(T)}}$}} and OH of clock $C_{B}$ at LTS $H_{_{LTS}}$ are  given in Table \ref{tab:4}. The  results denoted as  ``preprocessed'' and ``EEMD''  are obtained by taking all measurements in Period 2 as one unit. The  results denoted as  ``M-preprocessed'' and ``M-EEMD''  are obtained by implementing the grouping strategy for the measurements in Period 2 with 10-days measurement as different units, and then we  take the mean average of all units as the corresponding final results. The  results denoted as  ``W-preprocessed'' and ``W-EEMD''  are obtained by also implementing the grouping strategy for the measurements in Period 2 with 10-days measurement as different units, and we take the weighted average of all units as the corresponding results. Here we assign weights based on MDEV of each unit, which means that the higher the stability the greater the weight. Due to the fact that the frequency stability of the two hydrogen clocks used in experiments are at the level of $10^{-15}$/day, the accuracy of the determined geopotential is limited to tens of meters in equivalent height.

Compared to the corresponding EGM2008 model results which suggests that $ H_{LTS}=35.51$ m, the clock-comparison-determined results $H_{_LTS}^{_{(T)}}=133.2\pm43.5$ m. The deviation between the clock-comparison-determined results and the model value is 97.7 m. The results are consistent with the frequency stability of the hydrogen atomic clocks used in our experiments.

\renewcommand{\multirowsetup}{\centering}
\begin{table}
\caption{The clock-comparison-determined geopotential difference {\footnotesize{$\Delta W_{_{AB}}^{_{(T)}}$}} and OH of clock $C_{B}$ at LTS.  {\footnotesize{$D=H_{_{LTS}}^{_{(T)}}-H_{_{LTS}}$}} is the difference between $H_{_{LTS}}^{_{(T)}}$ and the EGM2008 results $H_{_{LTS}}$ ($H_{_{LTS}}=35.51$m).}
\label{tab:4}
\begin{tabular}{lllll}
	\hline\noalign{\smallskip}
	Time span & Strategy  & $\Delta W_{_{AB}}^{_{(T)}} (m^{2}/s^{2})$ & $H_{_{LTS}}^{_{(T)}}$ (m) & D(m)\\
	\noalign{\smallskip}\hline\noalign{\smallskip}
	\multirow{6}{3cm}{MJD \\*(58150-58215)} & &  & & \\
	&Preprocessed&$1987.5\pm2247.3$  &$254.4\pm229.3$ & $218.8\pm229.3$\\
	&EEMD &$1540.1\pm550.3$&$208.7\pm56.1$ &$173.2\pm56.1$\\
	&M-preprocessed &$2231.4\pm2192.2$&$279.3\pm223.7$& $243.7\pm223.7$ \\
	&M-EEMD &$1537.7\pm540.5$&$208.5\pm55.2$&$172.9\pm55.2$\\
	&W-preprocessed &$1875.4\pm2035.6$&$242.9\pm207.7$&$207.4\pm207.7$\\
	&W-EEMD&$ 1561.9\pm484.2$&$210.9\pm49.4$&$175.4\pm49.4$\\
	\noalign{\smallskip}\hline\noalign{\smallskip}
	\multirow{6}{3cm}{MJD \\*(58162-58215)} & &  & & \\    	
	&Preprocessed& $1424.9\pm2242.7$&	$197.0\pm228.9$&	$161.4\pm228.9$\\
	&EEMD           &$925.4\pm483.0$&$145.9\pm49.3$ &$110.5\pm49.3$ \\
	&M-preprocessed   &$1639.1\pm2202.5$&$218.8\pm224.7$ & $183.3\pm224.7$ \\
	&M-EEMD               &$932.7\pm483.7$ &$146.7\pm49.4$&$111.2\pm49.4$\\
	&W-preprocessed   &$1203.9\pm2038.5$&$174.4\pm208.0$&$138.9\pm208.0$\\
	&W-EEMD               &$ 799.9\pm426.0$&$133.2\pm43.5$&$97.7\pm43.5$\\
	\hline      	
\end{tabular}
\end{table}

\section{Conclusions }
\label{sec:7}
In this study, the clock comparison based on CVSTT technique for determining the geopotential difference is investigated. Based on this approach, one can directly determine the geopotential difference between two ground points. The experimental results provided in Table \ref{tab:4} show that the discrepancy between the clock-comparison-determined result and the corresponding model value is $97.7\pm43.5$ m. This is consistent with the frequency stability of the transportable H-master clocks (iMaser3000 and BM2102-02) used in our experiments.

Here we first used the EEMD technique to remove the periodic signals from the original CVSTT observations to more effectively extract the geopotential-related signals from clock offsets signals. We made comparisons between the results using EEMD and the results without using the EEMD technique, and our study shows that the results using EEMD is better, providing the deviation between the measured OH of the station LTS and  the EGM2008 model one with $D=97.7\pm43.5$ m. We stress that the EEMD technique is applied only to the preprocessing of the observations. If we don't use it, we still obtain comparable results with $D=138.9\pm208.0$ m, but a little worse. In addition,  as preliminary experiments, we used about 85-days observations to determine the geopotential difference between two stations based on the CVSTT technique, which might be applied extensively in geodesy in the future.

In our present experiments, the residual errors caused by the ionosphere, troposphere, and Sagnac effects can be neglected, for their effects are far below the accuracy level of the hydrogen clocks. However, to achieve centimeter-level measurement accuracy, those influences should be taken into consideration. Namely, more precise correction models need to be considered. In addition, we should use the time-comparison equation accurate to $1/c^{4}$ level based on GRT \citep{Kopeikin2011Relativistic}.


With rapid development of time and frequency science and technology, the approach discussed in this study for determining the geopotential difference as well as the OH is prospective and thus, could be implemented as an alternative technique for establishing height datum networks, and has potential applications for unifying the WHS, if GRT holds. 
Our experimental results are preliminary, and further experiments and investigations are needed to improve the results with high accuracy. \\

\noindent {\bf Acknowledgement}.  We would like to express our sincere thanks to three anonymous reviewers, the responsible editor, and the Editor in Chief of the Journal of Geodesy, for their valuable comments and suggestions, which greatly improved the manuscript.  This study was supported by the NSFC (grant Nos. 41721003, 41804012, 41631072,  41874023, 41429401, 41574007) and Natural Science Foundation of Hubei Province of China (grant No. 2019CFB611). \\

\bibliography{mybibfile}

\end{document}